\documentclass[sigconf%
]{acmart}

\usepackage{url} %
\usepackage{xspace} %
\usepackage{graphicx} %
\usepackage{color} %
\usepackage{amsmath}
\usepackage{mathtools}
\usepackage{xcolor} %
\usepackage{flushend}
\usepackage{hyperref} %
\usepackage{cleveref} %
\usepackage{enumitem}

\renewcommand\footnotetextcopyrightpermission[1]{}
\setcopyright{none}

\newcommand{\abbrev}{DR.SGX\xspace}

\usepackage{tikz}
\usetikzlibrary{shapes}
\newcommand*\circled[1]{\tikz[baseline=(char.base)]{
		\node[shape=circle,draw,inner sep=0.5pt] (char) {\footnotesize{#1}};}}

\begin{document}
\title[\abbrev]{\abbrev: Automated and Adjustable Side-Channel Protection \\ for SGX using Data Location Randomization}

\author{Ferdinand Brasser}
\affiliation{\institution{Technische Universit\"at Darmstadt}}
\email{ferdinand.brasser@trust.tu-darmstadt.de}

\author{Srdjan Capkun}
\affiliation{\institution{ETH Zurich}}
\email{srdjan.capkun@inf.ethz.ch}

\author{Alexandra Dmitrienko}
\affiliation{\institution{University of W\"urzburg}}
\email{alexandra.dmitrienko@uni-wuerzburg.de}

\author{Tommaso Frassetto}
\affiliation{\institution{Technische Universit\"at Darmstadt}}
\email{tommaso.frassetto@trust.tu-darmstadt.de}

\author{Kari Kostiainen}
\affiliation{\institution{ETH Zurich}}
\email{kari.kostiainen@inf.ethz.ch}

\author{Ahmad-Reza Sadeghi}
\affiliation{\institution{Technische Universit\"at Darmstadt}}
\email{ahmad.sadeghi@trust.tu-darmstadt.de}

\settopmatter{printacmref=false, printccs=false, printfolios=true}
\fancyhead[]{}

\begin{abstract}

Recent research has demonstrated that Intel's SGX is vulnerable to software-based side-channel attacks.  In a common attack, the adversary monitors CPU caches to infer secret-dependent data accesses patterns. Known defenses have major limitations, as they require either error-prone developer assistance, incur extremely high runtime overhead, or prevent only specific attacks. 

In this paper, we propose data location randomization as a novel defense against side-channel attacks that target data access patterns. Our goal is to break the link between the memory observations by the adversary and the actual data accesses by the victim. We design and implement a compiler-based tool called \abbrev that instruments the enclave code, permuting data locations at fine granularity. To prevent correlation of repeated memory accesses we periodically re-randomize all enclave data. Our solution requires no developer assistance %
and strikes the balance between side-channel protection and performance based on an adjustable security parameter.

\end{abstract}

\sloppy

\maketitle

\section{Introduction} 
\label{sec:intro}

Intel Software Guard Extensions (SGX)~\cite{Cos2016, ISCA2015} enable execution of security-critical application code, called \emph{enclaves}, in isolation from the untrusted system software. %
SGX was designed to ensure confidentiality of enclave data and integrity of enclave execution and is used in a number of academic works~\cite{haven,graphene,vc3,KPM+16,CKL+17,CDE+09,jitguard,smpc-sgx,fastkitten,voiceguard}.
Recent research has, however, demonstrated that SGX isolation can be violated using software-based side-channel attacks. In SGX, memory management, including paging, is left to the untrusted OS \cite{Cos2016}. By monitoring page usage, the OS can learn coarse-grained enclave control flow or data access patterns~\cite{Xu2015, Bulck2017}. Enclave data can also be inferred by monitoring CPU caches that are shared between the enclave and the untrusted software, enabling more fine-grained information leakage~\cite{BMD+17b, SWG+17, MIE17, GES+17, GRB+18}. Such attacks can defeat one of the main benefits of SGX---the ability to compute over private data on an untrusted (cloud) platform. 

The problem of side-channel leakage has been studied extensively. 
Oblivious RAM (ORAM) \cite{stefanov-ccs13} and Oblivious Execution \cite{maas-ccs13, liu-csf13, liu2015ghostrider} are well-known defensive techniques.
Obfuscuro~\cite{obfuscuro} implements those techniques for SGX enclaves, hiding all access patterns.
The main drawback is an extremely high runtime overhead ($83 \times$ on average and up to $220 \times$).
Another common defense is manual code hardening that is typically used by developers of cryptographic algorithms to make their implementations side-channel resilient~\cite{BrGrSe2006}. This defense is not easily applicable to enclaves written by developers who are not security experts.
Recent research has also proposed SGX-specific defenses. T-SGX~\cite{t-sgx} and D{\'e}j{\'a} Vu~\cite{incognito2017} use the processor's transactional memory features to prevent attacks that interrupt the victim enclave repeatedly. Such features are available only in a subset of SGX processors and the defense only protects against attacks that leverage interrupts. Cloak~\cite{cloak} and Raccoon~\cite{RaLiTi2015} hide memory accesses to developer-annotated enclave data, but relying on the developer to mark all (possibly non-obvious) secret data correctly can be very error-prone. 
In summary, all known defenses either impose extremely high runtime overhead, rely on the developer, require functionality that is not available in all CPUs, or mitigate only specific side channels. 

\paragraph{Our goals and approach.} In this paper we focus on information leakage caused by \emph{data access} monitoring. Our goal is to provide an \emph{automated} tool that provides side-channel protection without developer assistance and enables an \emph{adjustable} trade-off between security and performance. 

We focus on data accesses, as they are the target of many recent SGX attacks~\cite{BMD+17b, SWG+17, MIE17, GES+17}. Preventing control flow leakage is also important, but an orthogonal problem to our work.
We build an automated tool because, similar to the development of other software, not all enclave developers are security experts and many would fail to correctly use solutions that require identification of potentially subtle sources of leakage for manual annotation.
Instead, our primary goal is to strike the balance between provided protection and performance. 
While our tool can be configured to prevent all the leakage, this would incur a prohibitive performance penalty for most applications. Instead, we aim to give a means to enclave developers to get the best possible protection for a given application and performance overhead.

The main idea of our approach is to randomize all data locations in the enclave's memory at fine granularity. The enclave generates a secret randomization key and based on that computes a permutation for every memory address. As a result, the adversary cannot map the observed (permuted) memory address to the actual address, regardless of the channel he uses to make observations~\cite{BMD+17b, SWG+17, MIE17, GES+17,GRB+18, SGB+18, Xu2015, Bulck2017}. Because all data is randomized without the need to understand its structure or semantics, we call our approach \emph{semantic-agnostic data randomization}. 

Randomization is a well-known hardening technique, but our approach is different from the existing solutions that randomize code by leveraging its known structure, such as functions or blocks. Due to the well-known difficulty of C/C++ code analysis and pointer tracking, no similar structure is available for data~\cite{bigelow2015timely}. Indeed, existing randomization tools like SGX-Shield~\cite{SGXShield2017} focus on randomizing the code and do not tackle the problem of data randomization. Thus, they cannot prevent attacks that exploit data accesses, such as~\cite{BMD+17b, SWG+17, MIE17, GES+17}.

\paragraph{Challenges and results.} Secure and practical realization of our approach imposes a number of technical challenges. The first challenge is secure and efficient permutation computation under adversarial monitoring. If the adversary is able to derive information from the process of address permutation, he can revert the randomization. The second challenge is efficiency --- computing a permutation for every data access is expensive and causes a high overhead. The third problem is information leakage through repeated memory accesses. Although an individual access is effectively
hidden from the adversary, repetitive access patterns may allow (permuted) address correlation and leakage, i.e., correlation attacks.

In this paper, we tackle the above mentioned challenges and design and implement a compiler-based tool called \abbrev (\emph{\textbf{D}ata Location \textbf{R}andomization for \textbf{SGX}}) that instruments enclave code at compile time such that all memory locations used to store enclave data (in the heap) are permuted at cache-line granularity during run time.
We realize the permutation securely using small-domain encryption~\cite{ffx} and leveraging the CPU's hardware acceleration units (AES-NI). 
To address correlation attacks, our tool allows periodic re-randomization of enclave data: more aggressive re-randomization rates hide repeated memory access patterns better at the cost of higher run-time overhead.

The basic runtime overhead of \abbrev is $4.36\times$ without re-randomization.
Using different re-randomization rates, we measured an overhead approximately between $5\times$ and $11\times$. We acknowledge that this is a significant performance penalty, but emphasize that our solution is at least one order of magnitude faster than complete ORAM schemes like Obfuscuro~\cite{obfuscuro}.  Additionally, we note that this overhead only applies to the SGX enclave, which handles just the security-critical part of an application.

Our security evaluation reveals that the protection provided by \abbrev depends on the target enclave.
Enclaves where predictable data access patterns, like initialization routines, are soon followed by secret-dependent data accesses, require aggressive re-randomization to prevent leakage, incurring higher overhead. In a corner case, our solution can prevent any leakage by re-randomizing enclave memory after every memory access, effectively functioning as an ORAM implementation.
However, enclaves where secret-dependent accesses do not happen (soon) after predictable accesses can be strongly protected with much lower overhead. 

\paragraph{Contributions.} This paper makes the following main contributions:

\begin{itemize}[noitemsep,nolistsep]
 \item \emph{Novel approach.} We propose a novel approach called semantic-agnostic data
randomization as a defense against side-channel attacks on SGX.  
 \item \emph{New tool.} We design and
implement a tool called \abbrev that instruments code to permute an enclave's data memory locations at cache-line granularity and re-randomize them repeatedly.
 \item \emph{Evaluation.} We evaluate the
performance of our system, analyze possible leakage, and show how previous attack targets can be protected.
\end{itemize}

The  paper is organized as follows: \Cref{sec:model} defines our problem. \Cref{sec:design} presents our approach and \Cref{sec:impl} details on our implementation. We evaluate \abbrev's performance in \Cref{sec:performance} and analyze its security in \Cref{sec:analysis}.  \Cref{sec:relwork} reviews related work, \Cref{sec:discussion} provides discussion and
\Cref{sec:conclusion} concludes the paper.  

\section{Problem Statement} \label{sec:model}

In this work we focus on systems that provide an isolated execution environment that is implemented as an execution mode of the main CPU.
In particular, the CPU's shared resources, like caches, are used by all execution modes of the CPU and thus are shared between isolation domains.
Our work is targeted towards Intel SGX, however, the same model also applies to other architectures like ARM TrustZone~\cite{trustzone} and SANCTUARY~\cite{sanctuary} or software-based isolation solutions~\cite{trustvisor}. 

\paragraph{Problem space.}
Side-channel attacks on software in general, and SGX in particular, come in many different forms.
Any kind of resource use that is influenced by the software's execution and can be observed by the adversary can serve as a side channel.
For instance, the use of electricity as well as effects thereof like electro-magnetic emission, or the use of shared CPU caches.
In this work we focus on \emph{software} side channels, i.e., such that are observable by a software program running on the target machine, precluding physical or hardware side-channel attacks.

In the realm of software side-channel attacks a number of distinct variants exist.
On one hand, different shared resources can be used as a side channel, like the different caches of the CPU, or the virtual memory management.
On the other hand, side-channel attacks can target different information, including sensitive access patterns to data as well as secret dependent code execution paths. 

In this work we focus on software attacks that target \emph{data accesses} and consider attacks aiming to infer the control flow of a program as an orthogonal problem. Our rationale is two-fold. First, many side-channel attacks on SGX have been based on data access patterns~\cite{BMD+17b,SWG+17,MIE17,GES+17}. Furthermore, our solution can be combined with protections against control flow leakage attacks, for example with the Zigzagger approach proposed by Lee et al.~\cite{LSG+17}. 

\paragraph{Adversary model.}
\label{sec:advmodel}

The adversary's goal is to extract sensitive information from an isolated execution environment (enclave)~\cite{batina2019hardware, brasser2018advances} through cache side-channel attacks (including CPU-internal caches like the translation look-aside buffer~\cite{GRB+18}) and/or paging side-channel attack~\cite{Xu2015,Bulck2017}. 
Sensitive data in this context are not limited to cryptographic keys, which are the ``classical'' targets of side-channel attacks.
Instead, sensitive data have to be seen much broader, for instance, when processing privacy-sensitive data in the cloud~\cite{BMD+17b}.

The adversary can freely configure and modify all software of the system, including privileged software like the operating system (OS).
He knows the initial memory layout of the enclave, i.e., the code and initial data of the enclave.
Furthermore, we assume that the adversary can initiate the enclave arbitrarily often.

However, the adversary cannot directly access the memory of the enclave.
The internal processor state (e.g., the CPU registers) is inaccessible to the adversary, in the event of an interrupt the state is securely stored in an isolated memory region.
The adversary cannot modify the code or initial data of the enclave, as enclave's integrity can be verified using remote attestation.

We consider our work orthogonal to the recently discovered platform vulnerabilities Meltdown~\cite{Lipp2018meltdown} and Spectre~\cite{Kocher2018spectre} that leverage transient execution to read secrets across isolation boundaries. Although these vulnerabilities apply to SGX enclaves as well~\cite{sgxpectre, foreshadow}, Intel 
has already issued security updates for SGX that address such attacks~\cite{sgxpectre}. Also, SGX platform keys from unpatched (and thus potentially compromised) platforms can be identified at the time of attestation and revoked~\cite{sgxpectre}. 
The more general problem of data-access driven side-channels is much harder to solve in architectures like SGX. \abbrev addresses this latter and more difficult problem.

We assume the position of the attacker to be as strong as possible and therefore we will assume him to have a noise-free cache side-channel and to be able to obtain a ``perfect cache trace'' of the enclave. 
This means that he can observe all memory accesses of an enclave, e.g., using a cache attack technique such as Prime+Probe~\cite{Osv2006}.
He can precisely determine which cache line has been used by the enclave and also the order in which the cache lines have been accessed.
The adversary cannot extract information which is more fine grained than accesses to cache lines, i.e., the offset inside a cache line is not observable to him (see \Cref{sec:subline} for a discussion of possible attacks with finer granularity).
Additionally, for each memory access, the adversary can gain information about the accessed memory pages of an enclave~\cite{Xu2015,Bulck2017}. 

More formally, trace $t = \{c_1, p_1\}, ..., \{c_n, p_n\}$ is an ordered list of side-channel observation  pairs that capture every memory access that the victim enclave makes. In each observation pair, $c_i$ is the part of the memory address that determines the cache line the accessed address gets mapped to and $p_i$ is the part of the address that determines the accessed memory page. On current Intel CPUs the cache line size is $64$~bytes, thus, the last six bits of an address are oblivious to the adversary.
 
\paragraph{Design goals.}
General statements about which memory accesses of a program could leak information are hard to make in practice. 
All memory accesses must be assumed to potentially leak information if the attacker can associate them with relevant data elements or structures.
For the adversary it is sufficient to distinguish two memory locations to learn one bit of information.
Those memory locations could be two different data structures, e.g., two variables, or different elements within the same data structure, e.g., different entries in a table.
To protect all possible programs, the data structures of a program and the elements within data structures both need to be randomized.

The goal of our work is to provide a protection mechanism against side-channel attacks that can be applied to \emph{arbitrary enclave programs without developer assistance}. 
In particular, the developer must not be required to follow any rules or guidelines for programming his application or add annotations to the source code.
While annotating ``critical'' data in general helps improving the performance of most solutions, it is also very error-prone: especially in non-cryptographic applications, it is not always obvious which accesses to data objects might leak sensitive information.
This is crucial as most software developers are not security experts and cannot comprehensively identify data that could leak information.

The goal of \abbrev is to provide a trade-off between security and cost
in the design space reaching from unprotected processes, over plain SGX enclaves, enclaves with \abbrev to oblivious RAM (ORAM) solutions.
On the one hand, plain SGX enclaves provide basic data protection with little performance penalty;
on the other hand, schemes like Obfuscuro~\cite{obfuscuro} that implement ORAM for every memory access impose very high performance overheads ($83 \times$ on average and up to $220 \times$).
\abbrev strives to protect enclaves better than plain SGX while keeping the performance overhead at least one order of magnitude lower than systems like Obfuscuro.
\abbrev's security parameter (the re-randomization window $w$: see Section~\ref{sec:design}) allows it to be configured to cover the spectrum between plain SGX and full ORAM for data accesses.
With $w = 1$ \abbrev implements ORAM, admittedly in a costly way. 
On the other hand, $w = \infty$ only randomizes the initial memory layout of an enclave, which can be sufficient for some enclaves; we discuss this scenario in \Cref{sec:analysis}.
For most enclaves a window size between those two extremes can be chosen.
We evaluate different windows sizes in \Cref{sec:performance}.

\section{\abbrev} \label{sec:design}

Our core idea is to break the link between side-channel observations made by an attacker and the sensitive information processed by the victim.
Side-channel attacks inherently rely on the correlation between an observable effect and the data the attacker aims to extract.
Our defense obfuscates the link between memory locations and data elements.
Data elements are located at randomized memory locations, so the adversary cannot deduce which data element was accessed from an observed memory access location.
The adversary no longer learns \emph{which} data element was accessed but only learns that \emph{some} data element was accessed.

\begin{figure}[t]
	\centering
	\includegraphics[width=.85\linewidth]{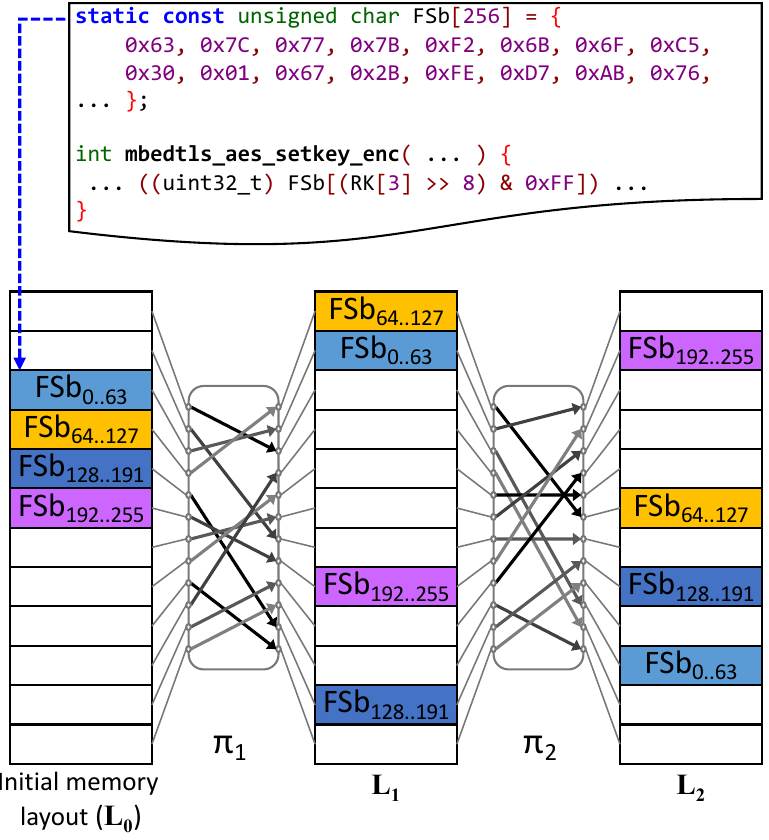}
	\caption{\abbrev's memory block randomization splits large memory structures like arrays into small blocks and reorders them. During the run time of an enclave its memory layout are re-randomized using the permutation function $\pi$. Each memory block is the size of a cache line ($64$~B), i.e., the finest granularity observable by the adversary.}
	\label{fig:idea}
\end{figure}

\abbrev splits enclave memory into small blocks that are randomly reordered, resulting in an unpredictable memory layout from the adversary's point of view.
\Cref{fig:idea} illustrates the concept on the example of the S-box of an AES implementation.
By default the S-box ($\texttt{FSb}$) is stored as an array in consecutive memory at a predictable location, shown on the left as initial memory layout $L_{0}$ in \Cref{fig:idea}.
Through a cache side channel an adversary can observe which part of the S-box is accessed.
Since the accesses to the S-box depend on the secret key the adversary can use this information to recover the key.
However, the adversary cannot observe accesses to individual bytes of the S-box but only at the granularity of cache lines ($64$~bytes).
\abbrev divides \emph{all} data memory of an enclave into blocks of cache line size, illustrated by the blocks forming $L_0$ in \Cref{fig:idea}.
These blocks are reordered by a permutation function $\pi_{1}$, resulting in a randomized memory layout $L_{1}$.
Throughout the runtime of an enclave the memory layout is constantly re-randomized, by applying a permutation function $\pi_{2}$ on $L_{1}$ a new and different memory layout $L_{2}$ is created.
As a result, the memory locations and thus the cache lines corresponding to the S-box are frequently changing, hindering the adversary's ability to link observed (cache or paging) accesses to the \mbox{S-box}. 

\subsection{Requirements and Challenges}
Below we describe the main challenges to tackle when implementing this idea.

\paragraph{Semantic gap.}
Providing side-channel protection through data randomization without developer assistance (e.g., code annotations) is a challenging task due to the semantic gap that is inherent to unsafe languages like C and C++.
Currently C and C++ are the only programming languages officially supported in the software development kit (SDK) that Intel provides for the development of SGX enclaves.

\paragraph{Re-randomization.} \label{sec:rerand}
Randomizing the memory layout of a program once to prevent an adversary from learning which data has been accessed is not sufficient.
The adversary can determine the relation of memory locations and data objects based on various information.
For instance, the initialization of data structures can reveal data locations.
In the example in \Cref{fig:idea}, the S-box is initialized during the creation of the enclave, however, other AES implementations initialize the S-box at run time which allows the adversary to learn the locations of all parts of the S-box array \emph{after} the initial randomization of the memory layout.
Similarly, access frequency can reveal the randomized location of data elements: if a particular object is accessed a predictable number of times the adversary can identify the object by finding the memory location that was accessed the expected numbers of times (frequency analysis).
To thwart the adversary in recovering the randomized memory location of data objects, their locations need to be changed throughout the runtime, such that the adversary cannot link data accesses to data objects.

\paragraph{(Re-)randomization under attacker's observation.}

All mem\-ory-related actions of the attacked enclave can be observed by the adversary, including those required during the initial data randomization and during the re-randomization of the memory layout.
The initial (un-randomized) memory layout is known to the adversary, i.e., he can monitor memory events while data is copied to its randomized locations.
Similarly, if the adversary managed to recover information about the randomized memory layout $L_n$ the adversary could link the re-randomization operations used to transfer data from $L_n$ to $L_{n+1}$ and thus also gain knowledge about the new layout $L_{n+1}$. 
Therefore, the randomization has to be done in such a way that its effects are not observable by the adversary.

\subsection{\abbrev Design}

\begin{figure}[t]
	\centering
	\includegraphics[width=.7305\linewidth]{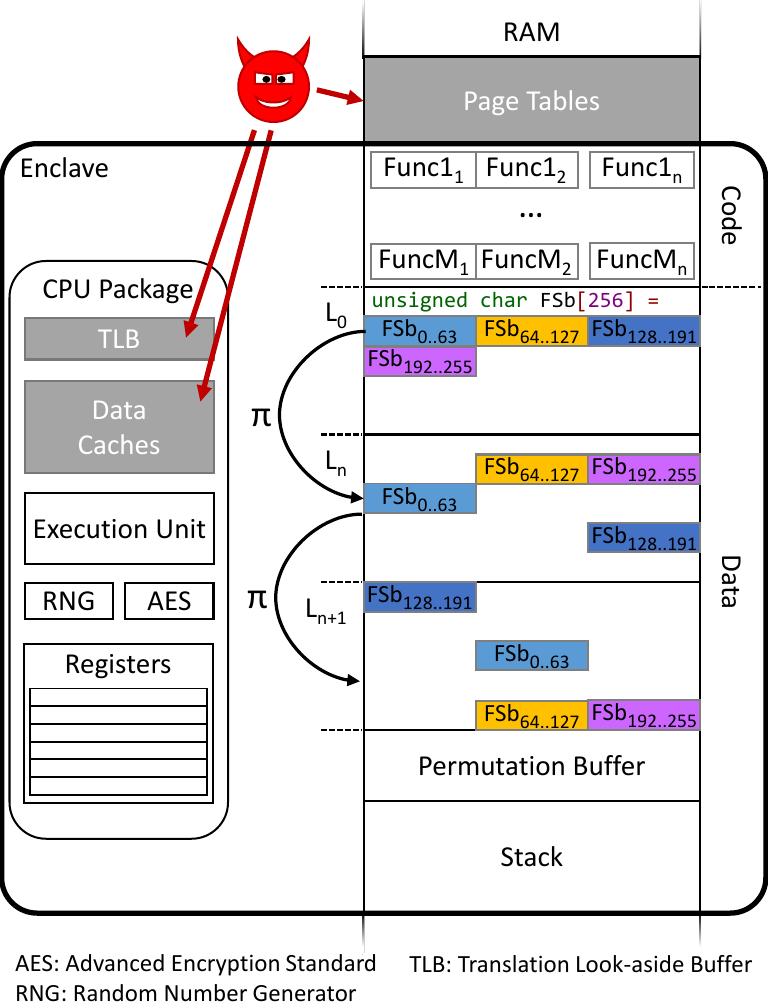}
	\caption{\abbrev's system design. The main memory of an enclave is not directly accessible by the adversary, however, the adversary can observe memory access indirectly through cache and paging side channels. The CPU's internal state stored in registers and/or special function units (e.g., the AES engine) are not observable by the adversary.}
	\label{fig:design}
\end{figure}

Our solution, a compiler-based tool called \abbrev, addresses the design goals and challenges described above by randomizing \emph{all} program data at fine granularity and re-randomizing the data continuously throughout the run time of the program. 

\Cref{fig:design} shows the system view of \abbrev.
The trusted computing base (TCB) of an SGX enclave includes the CPU package and an isolated section of the main memory (RAM).
However, the CPU caches, translation look-aside buffer (TLB) and the page tables are observable by the adversary.
The data cache of the CPU can be used to observe memory access patterns of an enclave.
On the other hand, the paging mechanism can be exploited in different ways to learn about memory reads and writes by an enclave.
By observing cache conflicts in the TLB, the adversary learns which memory pages were used.
Additionally, the adversary has control over the page tables also allowing him to learn which memory pages an enclave accessed.

However, an SGX enclave also includes components that cannot be attacked through a software side channel.
The CPU's registers and accesses to them cannot be observed by the adversary.\footnote{
	The LazyFP~\cite{LazyFP} attack cannot be used on SGX enclaves, since the register state is cleaned by the processor before exiting the enclave.}
Also the execution unit and special function units, like the random number generator (RNG) or the AES engine, are secure when operating over registers.
\abbrev combines these parts and function units of SGX that are secure against side channels, to obfuscate main memory accesses to the adversary.

\abbrev performs randomization at granularity of cache lines, the finest granularity at which the adversary can distinguish memory accesses (\Cref{sec:model}).
\Cref{fig:design} shows how \abbrev uses a random permutation function $\pi$ to reorder the program's data in memory.
Since the adversary cannot identify individual elements within a single cache line, accesses to the first array element ($\texttt{FSb[0]}$) and the $64$th element ($\texttt{FSb[63]}$) are indistinguishable for the adversary.
The randomization is based on secret values which are generated and only accessible \emph{inside} the enclave and only processed by the hardware AES engine of the CPU.
The CPU's AES engine holds all state and intermediate results in registers which are not observable by the adversary, hence, the adversary cannot learn about $\pi$ through cache or paging side channels.

\abbrev randomizes global variables and the heap. The stack cannot be easily randomized, since the hardware expects it to be contiguous.
Thus, variables on the stack larger than a cache line are moved to the heap, and replaced by a pointer on the stack.
The remaining variables are protected using multiple memory layouts:
for every function $n$ variants are created ($\mathtt{Func1_1}$, $\mathtt{Func1_2}$, ..., $\mathtt{FuncM_n}$ in \Cref{fig:design}), all with different stack memory layouts.
On every invocation of a function one of its $n$ variants is chosen randomly.

The size of the memory region (heap) for the enclave's data is a parameter of the permutation function $\pi$ (see \Cref{sec:impl}).

\paragraph{Memory access instrumentation.}
\abbrev performs randomization on cache line granularity for two reasons: (a)~randomizing at finer granularity provides no security advantages, and (b)~randomizing in a data structures aware fashion is impractical due to the semantic gap.
Our randomization requires that \emph{all} memory accesses are instrumented, which we ensure using a compiler pass.
The program code determines the memory location (i.e., address) of the data in the original, un-randomized layout.
Then, before the access is performed, the randomized location of that address is calculated.
The data is then accessed in its new, randomized location.

As we will elaborate in later sections, the cost of performing the randomization calculation for \emph{every} memory access is significant.
We overcome this problem by implementing a ``permutation buffer''.
The permutation buffer, similar to an address translation cache, holds the randomized locations of recently used data.
Hence, for data locations stored in the permutation buffer the function $\pi$ does not need to be recalculated.
However, accesses to the permutation buffer itself must be protected from leaking information.
Therefore the buffer is accessed in an oblivious way.

\paragraph{Initial randomization.} \label{sec:init}
The initial randomization of the enclave's data needs to be done in a way that cannot be observed by the adversary, to keep him from learning the randomization function $\pi$ or the new memory layout.
In particular, if the adversary can observe a read operation from the un-randomized initial memory layout and a subsequent write operation to a randomized address, he can link data structures to the randomized memory locations.

A general approach to break this linkage is to load a set of data into CPU registers (register operations cannot be tracked by the adversary) and write the data in a random fashion to their new locations.
This approach, however, is limited in the amount of the data that can be loaded at once into registers, enabling the adversary to learn partial information about the randomized memory layout.

\abbrev uses a randomization method which hides fine-grained (cache-line granularity) memory locations from the adversary. Specifically, we use \emph{non-temporal writes}~\cite{Intel-manual} that evade the CPU's caches, therefore the adversary cannot observe memory addresses written during the initial randomization.
Although the non-temporal writes prevent accesses to the new memory layout $L_{1}$ from being cached, the adversary can still observe the written memory locations through the more coarse-grained paging side-channel (that is, the adversary's trace contains a page event $p_i$, but no cache event $c_i$ for the non-temporal write).
This allows him to know, for each memory block read from the previous memory layout $L_0$, to which memory page it was written in $L_{1}$.
However, multiple cache lines are written to each page: assuming $4$~KB pages, $64$ cache-line-sized memory blocks will be written to the same page.
To hide this access pattern the initial randomization of \abbrev accesses \emph{all} memory pages of $L_1$ for each memory block that is moved, see \Cref{sec:without-predictable}.

\abbrev continuously re-randomizes the memory layout.
Starting from the initial memory layout $L_0$ a random permutation function $\pi_1$ is applied to derive the first randomized layout $L_{1} = \pi_1 \big( L_0 \big)$.
After a configurable window $w$ the memory layout is re-randomized, applying $\pi_2$ to derive $L_2 = \pi_2 \big( L_1 \big)$.

Like with the initial randomization, the adversary (who can observe reads from $L_{n}$ and writes to $L_{n+1}$) could link those operations to learn the relation between those memory layouts.
Again, \abbrev uses non-temporal writes to hide this information.
In \Cref{sec:analysis} we explain how a small number of re-randomization rounds hides the location of the element from the adversary completely.

\section{\abbrev Implementation} \label{sec:impl}

This section provides further details of \abbrev. We explain how we implemented the key-components of \abbrev: access instrumentation, permutation computation, initial randomization, permutation buffering, and 
re-randomization.
Throughout this section we will refer to \emph{data} memory regions or \emph{data} memory accesses simply as memory regions and accesses (omitting \emph{data}).

\begin{figure}[t]
	\centering
	\includegraphics[width=\linewidth]{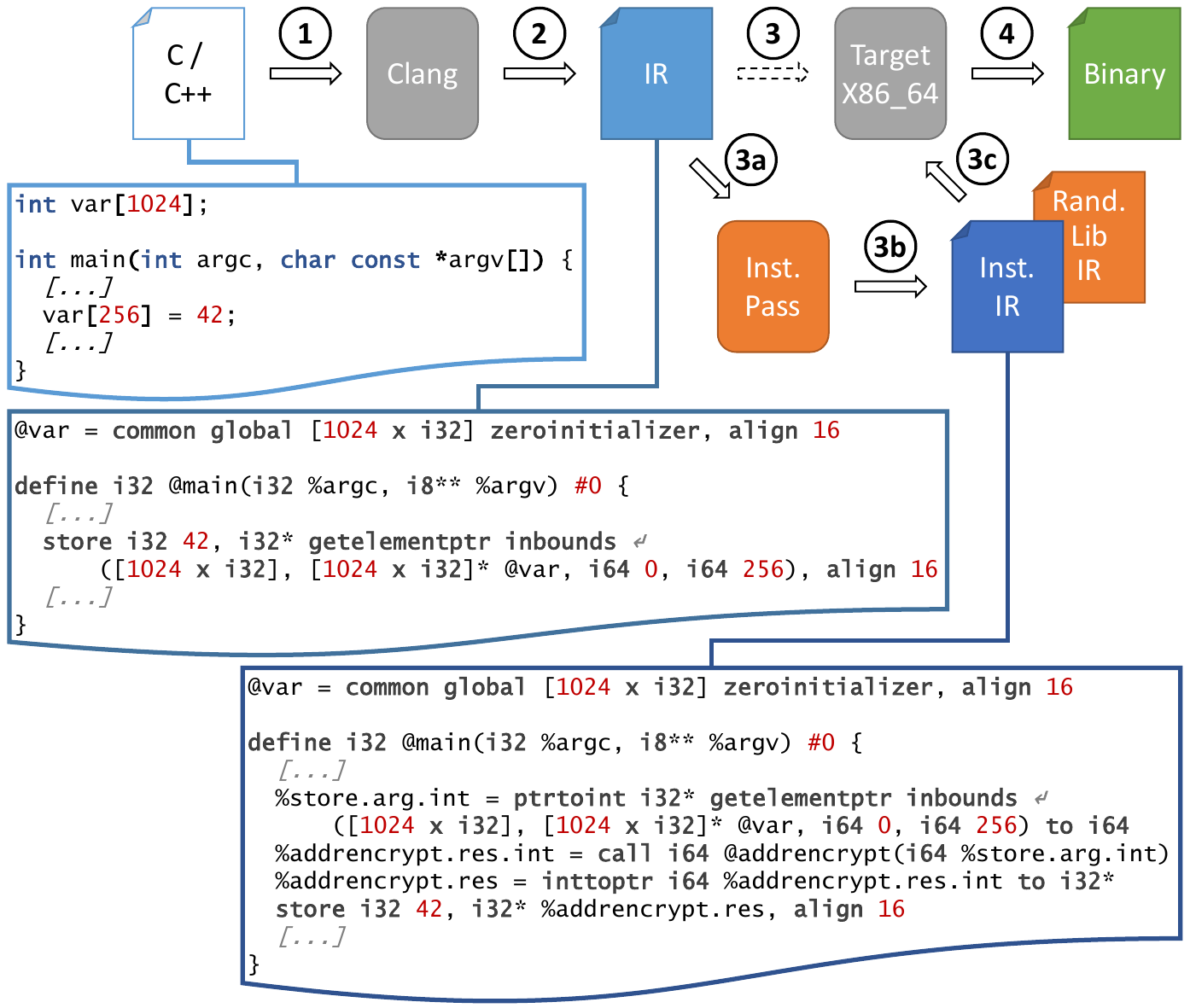}
	\caption{Code instrumentation with \abbrev. Before each memory access %
	the randomized memory address is calculated. The calculation is done by a function provided by the \abbrev library (Rand.\ lib), which can be written in C / C++ and is included in the instrumented binary. The snippets show the instrumentation of a sample store instruction.}
	\label{fig:instrumentation}
\end{figure}

\subsection{Memory Access Instrumentation}

\abbrev randomizes the memory locations of an SGX enclave's data.
The enclave, however, has been developed targeting a linear (virtual) memory model.
Therefore, each memory access of an enclave has to be instrumented to determine the correct randomized memory location of the data element that is meant to be accessed.

We extended the LLVM compiler~\cite{llvm} to instrument the enclave code, working at the intermediate representation (IR) level.
\Cref{fig:instrumentation} shows on the top the high-level compile process of LLVM.
A source file on the left is translated by the compiler front-end \circled{1}, Clang in the case of C/C++, into a LLVM intermediate representation (IR) \circled{2}.
The IR is then translated by the back-end \circled{3} into target architecture specific binary code \circled{4}, which in our case is Intel x86 64-bit.
With \abbrev the IR file is processed by a compiler pass \circled{3a} that instruments all memory access instructions (instrumentation pass) before it is translated into machine code \circled{3c}.
Furthermore, \abbrev adds a small library \circled{3b}, which contains functions used to perform the randomization.
This library can be written in a high-level language like C/C++ and is translated into IR as well.

Additionally, the instrumentation pass examines all allocations on the stack and transforms those which are larger than a single cache line into heap allocations.
A pointer to the heap allocation is placed on the stack and the code is modified to access the heap allocation instead of accessing the stack.

\paragraph{Instrumentation example.}
\Cref{fig:instrumentation} illustrates the instrumentation of a write access to an array.
The code snippet in the C file shows a write access to the $257$-th element of an integer array $\mathtt{var}$.
The code snippet in the middle shows the intermediate representation (IR) of the write operation.
The array is accessed by calculating the pointer to the $257$-th element of the array, using the LLVM function $\mathtt{getelementptr}$.
The value $42$ is then stored into this memory location.
The instrumented IR is shown in the bottom code snippet.
Again, a pointer to the $257$-th element of $\mathtt{var}$ obtained using $\mathtt{getelementptr}$ and stored in the variable $\mathtt{store.arg.int}$.
However, before storing the value $42$, $\mathtt{store.arg.int}$ is passed to the permutation function $\mathtt{addrencrypt}$.
The function returns the permuted location of the $257$-th element of $\mathtt{var}$, which gets cast from an integer value to a pointer value ($\mathtt{inttoptr}$).
The value $42$ is then stored to the permuted location $\mathtt{addrencrypt.res}$.

\subsection{Random Permutation}
\abbrev uses run-time data randomization, which is required for both the unobservable initial randomization as well as the re-randomizations.
This means that the randomized location of data must be recovered dynamically.
Using a purely random permutation would require storing extensive meta-data, which would then need to be accessed in an unobservable way.\footnote{The need to maintain meta-data is one of the main problems when using ORAM to protect SGX enclaves from side-channel attacks targeting the enclave's main memory accesses.}
Therefore, \abbrev uses a pseudo-random permutation function to determine the random location of data.
This approach has two advantages: (1) collisions, i.e., different element mapped to the same location, are inherently avoided, and (2) randomized locations can be computed based on a non-secret algorithm and a key, which is \emph{small} compared to the meta-data in the naive approach.
However, the permutation function itself must be resilient against side-channel attacks, otherwise the adversary can learn the randomization secret and disclose the accessed memory locations.

We use small-domain encryption for our random permutation function. The domain size must be in the order of memory size used by the enclave employing \abbrev (divided by the size of a cache line).
In particular, we use the FFX Format-Preserving Encryption scheme, which is based on a $10$-round Feistel network~\cite{ffx}.
As the underlying block cipher for FFX we used AES, for which the hardware acceleration extension AES-NI~\cite{Intel-manual} is available in all SGX-enabled CPUs.
AES-NI provides both good performance and resiliency against cache-based side-channel attacks.

Our implementation only supports single-threaded enclaves. However, standard software-engineering techniques can be employed to extend the support to multi-threaded enclaves. 
Only the re-randomization operations need to be synchronized between threads.

\subsection{Initial Randomization}
\label{sec:impl:initrand}

The initial randomization is particularly challenging since the adversary knows the initial memory layout of an enclave.
If we used standard write operations to copy data from the initial data section $L_0$ to the randomized section $L_1$, the adversary would be able to learn the randomized layout.

In \abbrev we use non-temporal write instructions to tackle this problem~\cite{Intel-manual}.
Non-temporal write instructions provide the processor with the meta-information that the data will not be used again soon by the program and it is not necessary to store them in the cache.
On current Intel processors memory write operations using this instruction immediately affect the DRAM and are not buffered in the CPU's cache,\footnote{We verified this behavior on a Skylake test system by issuing a non-temporal write followed by a read from the same cache line, and verifying that the read generates a cache miss on all three cache levels.} i.e., they are invisible to the adversary.
Page-granularity side-channels information is hidden by accessing all heap memory pages for each block.

The secret keys we need as input to our random permutation are generated by the hardware random number generator \emph{inside} the enclave.
We use $\mathtt{rdseed}$ to obtain true random numbers from the CPU~\cite{Intel-manual}.
This way the adversary cannot influence or obtain the secret key. 

\subsection{Stack Randomization} \label{sec:impl:stack}
\abbrev uses the stack only for data elements that are smaller than a cache line, all other data are moved to the heap where they are subject to (re-)randomization.
For the remaining data elements on the stack we use an approach inspired by the code randomization method introduced by Crane et al.~\cite{CHB+15}.
The stack layout of each function is randomized by reordering the local variables on the stack.
At compile time $n$ variants of each function with different stack layouts are generated. 
At run time one function variant is chosen at random every time it is invoked.
\abbrev uses $n=10$ variants for each function, as the empirical evaluation~\cite{CHB+15} suggests.

\subsection{Permutation Buffer}
\label{sec:permubuffer}
Performing the calculation for the pseudo-random permutations is costly and needs to be performed for each memory access.
To improve the performance we introduced a buffer for memory translations (Permutation Buffer in \Cref{fig:design}).
Permutation is performed at cache line granularity, i.e., all bytes in one cache line in $L_0$ are mapped as a single block.
When this block is moved to $L_1$ it will, with high probability, be mapping to a different cache line, and to yet another cache line in $L_2$, and so on.
On recent x86 processors a cache line is $64$~bytes, thus, by storing the result no extra calculations are necessary for memory accesses that fall within the same cache line.
Our buffer is currently $1$~KB which allows for a direct-mapped storage of permutation results for $256$~translations.
To prevent leakage through our permutation buffer we access it in a way which is oblivious to the adversary. 
For each read operation to the buffer we simply access \emph{all} CPU cache lines in our permutation buffer.
Moreover, we randomize the location of the items in the permutation buffer by performing an xor operation with a randomly-generated value before determining which buffer item to use.
The random value changes and the buffer is invalidated every time a re-randomization happens.

\subsection{Re-Randomization}

\abbrev constantly re-randomizes the memory layout of an enclave.
\Cref{fig:design} shows the overall memory layout.
The blocks are copied from $L_n$ to $L_{n+1}$ in the same order as they appear in $L_n$, so the adversary only observes reads to every block in $L_n$, in order.
Like in the initial permutation, non-temporal write operations are used to hide fine-grained writes. 

For each cache-line-sized memory block in $L_n$, \abbrev needs to compute the corresponding addresses in $L_n$ and in $L_{n+1}$.
Hence, the cost of re-randomization primarily comes from the permutation calculations required.
However, the pipelining of AES instructions in the CPU makes encrypting multiple addresses together faster than encrypting them sequentially.
This reduces the cost for the re-randomization and leads to better overall performance of \abbrev.

\section{Performance Evaluation} 
\label{sec:performance}

We evaluated the performance of \abbrev using the benchmark suite Nbench~\cite{nbench}.\footnote{Benchmarking SGX code can be challenging, since well-known benchmark suites rely on a number of features, including system calls, timestamps, and the file system, which are not directly available in SGX.}
We use Nbench because it has been previously used to analyze SGX performance~\cite{SGXShield2017},
it relies only marginally on the file system, and it is relatively simple (5217~LoC), so it can easily be adapted to run inside an SGX enclave.
The original version relies on timestamps to run each benchmark for an equal amount of time; since timestamps are not available in SGX enclaves we manually chose for each benchmark the lowest number of iterations that yielded a run time greater than 100~ms.
We measured the run time of the benchmarks by briefly switching to the non-SGX mode and reading the hardware time stamp counter.
We measured the overhead due to this mode switch and it is negligible compared to the overall run time.
Our test system is equipped with an Intel Skylake i7-6700 processor clocked at $3.40$~GHz, $128$~MB Enclave Page Cache, running Ubuntu 14.04.4.

\paragraph{Memory overhead.} \label{sec:performance:memory}
The memory overhead of \abbrev is mainly due to (1) heap randomization and (2) stack randomization. For the heap randomization two memory areas as large as the heap need to be reserved while the re-randomization is in progress. In our evaluations the heap size was set to values between $512$~KB and $4$~MB. Whenever the re-randomization is ongoing an additional $100\%$ for the heap size is required.
Stack randomization is based on providing $n$ variants for each function. This increases the memory required for the code by factor $n$. We chose $n=10$, thus the overhead is $10\times$. The size of the stack itself does not increase. For each invocation at run time only one of the function variants is used, i.e., the number of stack frames to be stored on the stack does not increase.

\begin{figure}[tb]
	\centering
	\includegraphics[width=\columnwidth]{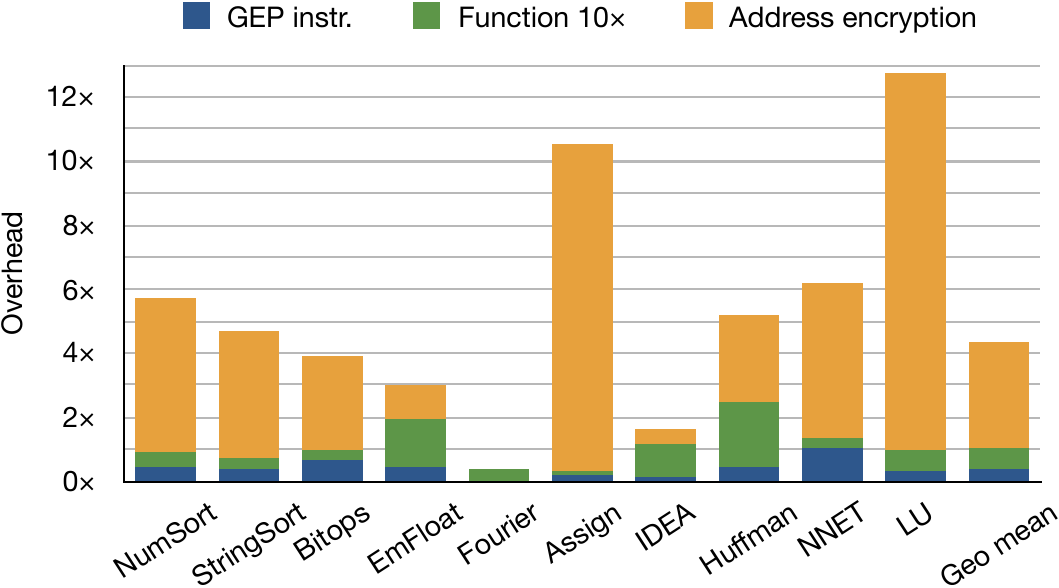}
	\caption{Overhead of each benchmark, using various subsets of \abbrev.}
	\label{fig:pe-part}
\end{figure}

\paragraph{Runtime overhead of \abbrev modifications.}
We evaluated \abbrev with different subsets of its components active. To get a better understanding of the impact \abbrev's individual components we ran all benchmarks multiple times activating one more components for every repetition.
A breakdown of each component's overhead is shown in \Cref{fig:pe-part}.

We first tested our mechanism to move large stack allocations to the heap, i.e., replacing allocations on the stack larger than $63$~bytes with calls to \texttt{malloc}.
We measured a negligible overhead well below $1\%$, which is too small to be visible in \Cref{fig:pe-part}.
Then, we tested the instrumentation of reads and writes (LLVM instruction \texttt{getelementptr}).
In \abbrev, instances of this instruction are followed by a call to our permutation function, unless the argument to the instruction is on the stack.
In this test, the identity permutation function was used, which returns immediately. Therefore overhead reflects the impact of the instrumentation alone.
We measured overheads between 0 and ${102}\%$, with a geometric mean of ${39}\%$ (\emph{GEP instr.} in \Cref{fig:pe-part}).
Next, we added our stack randomization using function duplication. The geometric mean of the additional overhead is $46\%$, while the maximum is $135\%$ (\emph{Function 10$\times$} in \Cref{fig:pe-part}).
Finally, we tested our complete system (without periodic re-randomization).
Overheads range between $0.39\times$ and $12.77\times$, with a geometric mean of $4.36\times$.
The benchmarks \emph{Assign} and \emph{LU} have the biggest overheads, $10.56\times$ and $12.77\times$ respectively, due to high miss rates in our permutation buffer (their miss rates are $\sim 13\times$ higher).

\begin{figure}[tb]
	\centering
	\includegraphics[width=\columnwidth]{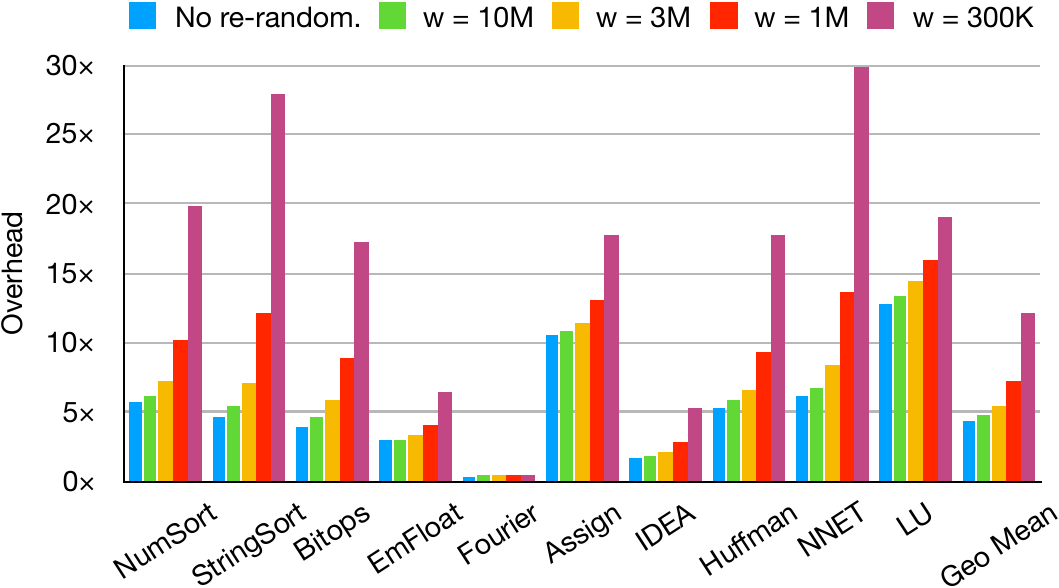}
	\caption{Overhead of each benchmark, with heap size $h$ = 4~MB, without re-randomization and with various re-randomization windows $w$.}
	\label{fig:pe-4MB-ovh}
\end{figure}

\paragraph{Runtime overhead of re-randomization.}
Next, we assessed the impact of various window sizes $w$ and heap sizes $h$ on the run time overhead and re-randomization window duration.
We chose our heap size $h \in \lbrace 4\text{~MB}, 2\text{~MB}, 1\text{~MB}, 512\text{~KB} \rbrace$ but other values are also possible.
We measured the time required to perform a re-randomization by dividing the CPU cycles required by the processor's nominal speed, $3.4\text{~GHz}$.
The re-randomization requires $7.31\text{~ms}$, $4.07\text{~ms}$, $2.26\text{~ms}$, and $1.26\text{~ms}$ respectively for $h=4\text{~MB}$, $h=2\text{~MB}$, $h=1\text{~MB}$, $h=512\text{~KB}$.

We first measured the run time overheads for $h = 4\text{~MB}$, $w \in \lbrace 10\text{~M}, 3\text{~M}, 1\text{~M}, 300\text{~K} \rbrace$.
In \Cref{fig:pe-4MB-ovh}, the left-most bars in each group represent the overhead without re-randomization (like \Cref{fig:pe-part}), with a geometric mean of $4.36\times$.
Re-randomization every 10 million accesses ($w=10\text{~M}$) increases the overhead slightly (geometric mean of $4.76\times$).
Reducing the window to $3\text{~M}$, $1\text{~M}$, and $300\text{~K}$ brings the geometric mean of the overhead to $5.45\times$, $7.29\times$, and $12.21\times$.

\begin{figure}[t]
	\centering
	\includegraphics[width=\columnwidth]{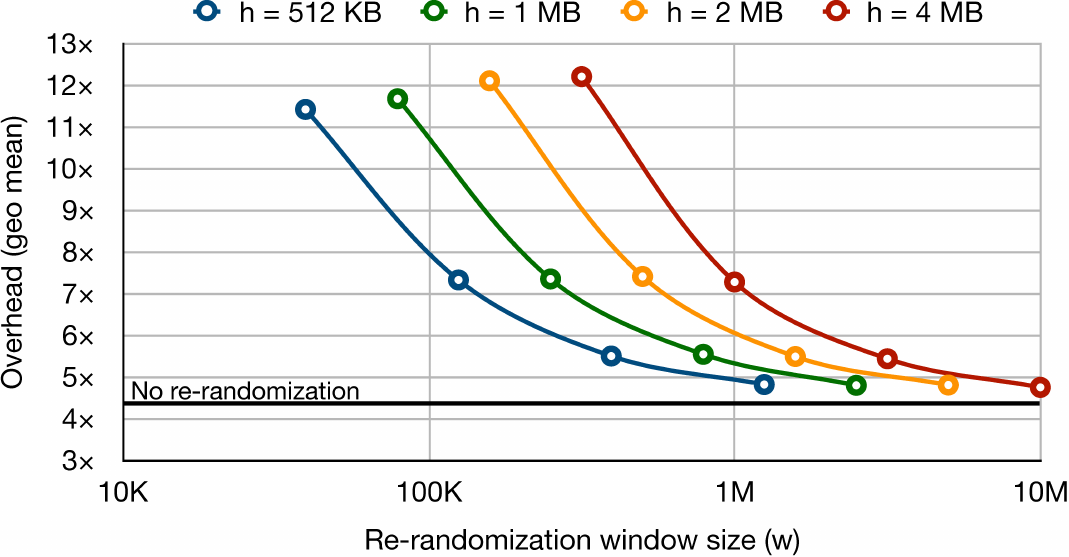}
	\caption{Geometric mean of the overheads, for various heap sizes $h$ and various window sizes $w$. The black line represents the overhead without re-randomization.}
	\label{fig:pe-wo}
\end{figure}

We then measured the overhead for smaller heap sizes.
We expected that halving both the heap size and the window size, i.e., re-randomizing a heap half as big twice as often, would yield similar performance results.
\Cref{fig:pe-wo} shows the overhead depending on the window size for various heap sizes and confirms our intuition.
Each line refers to a heap size twice as big as the line to its left.
The black line at $4.26\times$ is the overhead measured in the case without re-randomization and represents $\lim_{w\to\infty}$ of the overhead; in other words, increasing the values of $w$ further would bring diminishing results.

\paragraph{Summary.} The performance of our solution depends heavily on the user parameters. For example, the overhead is $4.8\times$ for parameters $h=\text{1MB}$ and $w=2.5\text{~M}$. 

Developers and system administrators can adjust the parameters of \abbrev based on the memory needs of their application and the available computing resources. For example, if the deployment scenario requires $1\text{~MB}$ of heap memory and allows up to $8\times$ overhead, the window size $w$ can be set to $250\text{~K}$ for maximal re-randomization rate and security (see \Cref{fig:pe-wo}). We consider this task of parameter tuning feasible for most developers. A typical developer may not be able to assess subtle sources of information leakage for correct source code annotation, but usually the developer knows the application's performance requirements and can set $h$ and $w$ accordingly.

Finally, we emphasize that in many SGX application scenarios, the overhead of the enclave (imposed by \abbrev) is not directly the overhead of the entire application. For example, SGX-based applications that perform networking or database queries spend most of their time in the unprotected part of the application, and therefore the slowdown of the enclave represents only a minor part of the application's performance. Thus, in many cases, a high enclave overhead can still be acceptable for the overall performance.

\section{Security Analysis} 
\label{sec:analysis}

In this section we analyze the security of \abbrev. We focus on the security properties of our novel heap data protection mechanism. Our stack data protection follows a known approach, evaluated in~\cite{CHB+15}. 

The goal of the adversary is to recover secret data from the victim enclave based on secret-dependent (heap) data access patterns to data. Recall that we consider a powerful adversary that gets a perfect trace of all cache and page events. Since all known attacks~\cite{BMD+17b,SWG+17,MIE17,GES+17} exhibit significant noise in the cache channel, this is an over-approximation of the capabilities of today's attackers and allows us to reason about the effectiveness of our solution against more powerful future adversaries.

\begin{figure*}[t]
	\centering
	\includegraphics[width=\linewidth]{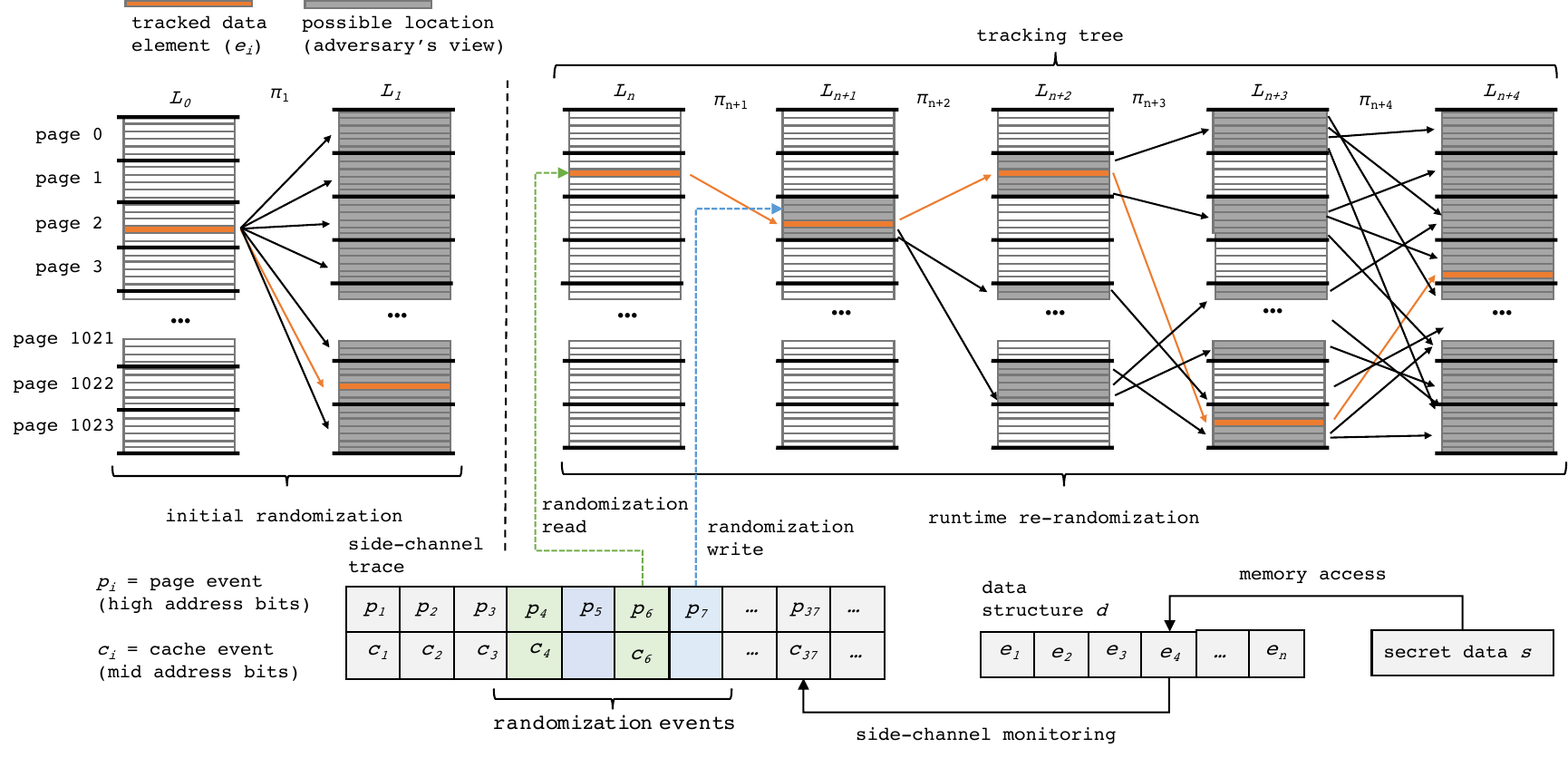}
	\caption{\textbf{Location tracking.} By identifying missing cache events in the trace, the adversary can learn the re-randomization writes (marked in blue) and preceding re-randomization reads (marked in green). 
	The initial randomization hides destination addresses completely.
	The adversary can build a tracking tree, where the source address of each re-randomization is known with cache-line granularity and the destination address with page granularity. After few re-randomization rounds, the tracked memory location can reside in any memory location.}
	\label{fig:analysis}
\end{figure*}

In a data-driven side-channel attack, the adversary leaks information by monitoring secret-dependent access patterns. We model this as follows. The targeted victim enclave has secret data $s$ of any length. The secret could be a cryptographic key, medical data, financial information or sensitive machine learning training sets. The enclave has a data structure $d$ that consists of $n$ elements ($e_1, ..., e_n$) and is accessed based on $s$. The data structure could be a look-up table, S-box, index, or in-memory database. The size of each element $e_i$ is the cache line size (smaller elements cannot be attacked, larger elements can be modeled as multiple elements). Based on the value of $s$, the enclave makes $k$ accesses to different elements of $d$. Such access pattern determines the value of $s$. The enclave may also make predictable accesses to $d$ (e.g., iterate through it during initialization).

\subsection{Finding Attack Position in Trace}

We start our analysis by explaining how the adversary can find the ``attack position'' in the side-channel trace, i.e., the position where (permuted) secret-dependent data accesses take place. 
The adversary can compile the victim enclave without \abbrev protection and instrument those parts of the enclave where the secret-dependent accesses to $d$ happen. The adversary can then run the instrumented enclave, monitor side-channels, and based on the instrumentation learn the position in the trace where the secret-dependent accesses are located. After that, the adversary can run the victim enclave that is protected with \abbrev using the same inputs and again monitor side-channels. Assuming a deterministic enclave,\footnote{We consider a deterministic enclave, because that is the best case the for adversary for building the tracking tree. Thus, the following analysis based on this assumption represents the best case for the adversary regarding finding the attack position in the trace.} the adversary obtains a protected trace that includes additional randomization events to the trace (see Figure~\ref{fig:analysis}). Next, the adversary can filter out all randomization events. Since we use non-temporal (NT) writes that bypass the cache for randomization writes, the adversary finds each page event $p_i$ that has no corresponding cache event $c_i$ in the trace. For each such randomization write, the previous event in the trace is a read due to the randomization. The adversary removes all randomization events. The known attack position in the non-protected trace corresponds to the same position in the filtered protected trace.

\subsection{Inferring Secret Enclave Data} 
\label{sec:without-predictable}

Once the attack position is known, the adversary can attempt to infer secret data $s$ from the permuted memory accesses in the attack trace. The adversary's success depends on the type of the victim enclave. 

\paragraph{No predictable accesses.}
We first consider enclaves that make \emph{no} predictable accesses to $d$ (i.e., the enclave accesses $d$ only based on a pattern that is derived from the secret data $s$). For such enclaves, \abbrev provides strong protection due to its initial randomization that is illustrated in Figure~\ref{fig:analysis}. The enclave's data is copied from the known, original memory layout $L_0$ to a new randomized memory layout $L_1$ in blocks of cache line size using NT writes. For each block, the initial randomization process performs one read access to the original memory layout and NT writes to all memory pages. Because NT writes hide the accessed address at cache-line granularity, the adversary gains no knowledge of the new location in $L_1$. The same process is repeated for every memory block and in the end the location of each block in $L_1$ is equally likely for the adversary.

By observing the permuted side-channel trace, the adversary may infer execution characteristics such as frequencies of accesses to the same memory address (e.g., address $a$ was accessed $x$ times). However, because permuted addresses $a$ can refer to any actually accessed addresses, such frequency analysis does not help the adversary to infer the secret data $s$, unless the enclave exhibits predictable access patterns which we discuss below.

Assuming no predictable access patterns, the best option for the adversary is a guessing attack. The adversary knows the permuted addresses of $k$ secret-dependent accesses. For each access, every address, and thus every data structure element $e_i$, is equally likely. After observing $k$ distinctive accesses to $n$ elements, the number of possible alternatives will be given by an arrangement of $k$ from $n$: $A_n^k = \frac{n!}{(n-k)!}$. For example, a data structure of $n=50$ elements and any number of secret-dependent accesses resulting in 25 distinctive accesses to the data structure, the amount of arrangements is $1.96 \times 10^{39}$, which gives the chance of a random guess of approximately $2^{-131}$. We conclude that \abbrev provides strong protection for enclaves that have no predictable accesses to the data structure $d$.

\paragraph{Predictable accesses.}
The second case that we consider is a victim enclave that exhibits predictable access patterns to $d$, e.g., the enclave may initialize $d$ in an order that is known to the adversary. The enclave may also access elements of $d$ a predictable number of times. Such predictable accesses in the trace will disclose the current permuted memory addresses for each accessed element $e_i$.

Figure~\ref{fig:analysis} illustrates an example scenario, where the permuted address of element $e_i$ is revealed to the adversary in memory layout $L_{n}$. The next re-randomization round moves the data of that element to a new location in layout $L_{n+1}$. Since the move operation is implemented using NT writes, the adversary learns the new page in $L_{n+1}$, but not the fine-grained location. The leakage of the target page allows the adversary to construct a \emph{tracking tree} for element $e_i$.

The expansion of the tracking tree depends on the size of the used memory in the victim enclave. For example, if the victim enclave uses 2 MB memory (out of total 4 MB address space), each memory page contains on the average 32 blocks. On the next re-randomization round, each of these blocks are moved to new memory locations in layout $L_{n+2}$. Because the adversary does not know the exact location of element $e_i$ in $L_{n+1}$, he cannot distinguish when the element is moved from the set of 32 move operations that use the same page as the source.\footnote{Our implementation randomizes 8 blocks at once which makes tracing even more difficult for the adversary.} From the adversary's point of view, after two re-randomization rounds, the element can reside in 32 pages with high probability. After four re-randomization rounds, the adversary must track $32^3 = 32,768$ re-randomization moves. Although some of the moves may write to the same target pages, the tracking tree covers all 1,024 memory pages in $L_{n+4}$ with high probability, and thus all memory locations are equally likely for the adversary. For enclaves with smaller heap size (512 KB), similar effect can be achieved after three rounds. The shortest re-randomization window we tested in \Cref{sec:performance} lasted 0.37 ms, in which case the required three (or four) re-randomization rounds would be performed after 1.1~ms (or 1.5~ms) of enclave execution.
We conclude that enclaves with predictable accesses can leak information. If the secret-dependent access happens after the predictable access and before a sufficient number of re-randomization rounds, the secret may be leaked to the adversary. By touching additional memory pages on every re-randomization write, the window can be reduced to fewer rounds. Alternatively, re-randomization rounds can be performed more frequently. Both approaches increase runtime overhead.

\section{Related Work} 
\label{sec:relwork}

Previous research has proposed various side-channel defenses. In this section we review them and compare existing defenses to \abbrev.

\paragraph{ORAM and Oblivious Execution}
Oblivious RAM (ORAM)~\cite{goldreich1987towards, GolOst1996, stefanov-ccs13, ren2015constants, goodrich2012privacy, williams2012round} refers to schemes that hide the memory access pattern of a trusted client (e.g., CPU or network client) to an untrusted and encrypted memory (e.g., DRAM or server) by introducing fake accesses and shuffling the encrypted memory elements such that the observable access pattern is independent of the actual access pattern.
Oblivious execution architectures~\cite{maas-ccs13, liu-csf13, liu2015ghostrider} attempt to hide all observable effects of program execution, including both memory accesses (code and data) and timing information.
Implementing ORAM for every enclave memory access is extremely expensive. Obfuscuro~\cite{obfuscuro}, a program obfuscation system, implements both ORAM and oblivious execution, with performance overheads of $83 \times$ on average and up to $220 \times$.
\abbrev's performance overhead is at least one order of magnitude lower than Obfuscuro.

Sinha~\cite{Sinha2017} proposes a compiler-based tool to protect code written in their custom language from paging-based side-channel attacks. In contrast, \abbrev works with existing code in C/C++ and also mitigates cache-based side-channel attacks.

Raccoon~\cite{RaLiTi2015} is a system that provides oblivious data access only for developer-annotated enclave data, thus reducing the overhead. Memory accesses are hidden by either using ORAM or by streaming over the entire data structure. In contrast, \abbrev does not rely on developers to identify and annotate data that might leak. 

ZeroTrace~\cite{sasyzerotrace} is an oblivious data structure framework for SGX that runs on top of a software memory controller. ZeroTrace is designed to hide memory access to resources \emph{outside} of an enclave, e.g., to the hard disk drive. Importantly, it is not designed to make \emph{all} memory accesses of an enclave to its own main memory oblivious, like \abbrev does. Furthermore, ZeroTrace requires the developer to use the memory controller interface for all access that should be protected. \abbrev does not require similar developer assistance.

Ohrimenko et.\ al.\ propose data-oblivious machine learning algorithms~\cite{OSF+16} and a side-channel resilient MapReduce framework~\cite{ohrimenko2015observing} for SGX.
Fuhry et.\ al.\ propose a page-fault side-channel secure database~\cite{hardidx}.
Such defenses are tailored to specific enclaves and algorithms, while \abbrev applies to arbitrary enclaves.

\paragraph{Transactional memory.} 
Some of the known SGX side-channel attacks interrupt the victim enclave repeatedly~\cite{Xu2015}. A corresponding defense is to enable the victim enclave to detect interruption and take counteractive measures, such as stopping its execution. T-SGX~\cite{t-sgx} leverages the Intel Transactional Synchronization Extension (TSX) to detect asynchronous enclave exits, e.g., due to interrupts of page faults.  D{\'e}j{\'a}~Vu~\cite{incognito2017} monitors the execution time of an enclave to detect a slowdown caused by frequent interrupts. These defenses do not prevent attacks that work without interrupts~\cite{BMD+17b, SWG+17, GES+17}. \abbrev is applicable to such attacks.

Cloak~\cite{cloak} uses TSX to preform atomic memory operations that hide sensitive memory accesses. Before sensitive memory is accessed, all cache lines are touched (primed) by the enclave, and thus the adversary learns nothing about the enclave's sensitive accesses. Cloak relies on the developer to annotate sensitive data structures that should be protected from side-channel attacks and requires TSX, which is not supported by all SGX processors. \abbrev does not require similar developer assistance and works on all SGX processors.

\paragraph{Software diversity.} Crane et al.~\cite{CHB+15} propose to apply dynamic software diversity, an effective countermeasure against code reuse attacks and reverse engineering, to defend against cache-based side-channel attacks. 
The approach is to create multiple copies of code and choose one of them at the time of execution. We apply this technique to protect stack data.
However, the solution by Crane et al.~is specifically targeting protection of cryptographic algorithms. In contrast, \abbrev can protect non-cryptographic enclaves.

\paragraph{Randomization.} Address Space Layout Randomization (ASLR)~\cite{PaX-ASLR} is a common defensive technique against memory corruption attacks such as ROP~\cite{roemer2012return}. ASLR hides the locations of memory regions (code and data) by randomizing their offsets at load time.  More fine-grained solutions randomize code (but not data) at function~\cite{kil2006address}, block~\cite{WaMoHaLi2012, TUD-CS-2013-0042}, or instruction~\cite{PaPoKe2012, Hi2012} level.

Such randomization techniques are insufficient as a side-channel defense for SGX. Offset-based ASLR is not effective since the privileged attacker is responsible for memory management and thus learns the ``secret'' randomized offsets. Code randomization, as implemented in SGX-Shield~\cite{SGXShield2017}, is not complete~\cite{guards-dilemma} and does not prevent attacks that monitor data accesses~\cite{SWG+17, Xu2015, BMD+17b}. 

\paragraph{New cache architectures.} Cache-based side channels can be addressed by changes in the cache architecture. The two common approaches are (i)~cache partitioning~\cite{domnitser2012non, Page05partitionedcache, WangLee2008, DoJaLoAbPo2012}, dividing the cache into partitions that are not shared between processes, and (ii)~cache access obfuscation~\cite{WangLee2008, Keramidas2008, LiLe2014, kong2009hardware, hybcache}, where the goal is to obfuscate the obtainable side-channel information, either by introducing noise or by randomizing the address to cache line mapping. 
Such defenses require hardware changes and are limited to cache attacks. \abbrev works on current processors and applies to additional side-channels (e.g., page faults).

\section{Discussion}
\label{sec:discussion}

\paragraph{Fine-grained leaks.}
\label{sec:subline}
Recent works~\cite{CacheBleed2016,MemJam} have investigated the possibility of leaking information through a side-channel with a granularity smaller than a cache line.
However, they are not applicable in our case.

CacheBleed~\cite{CacheBleed2016} exploits cache \emph{bank conflicts} to leak fine-grained information. 
This attack does not apply to SGX CPUs due to an updated cache design.
We verified this experimentally.

MemJam~\cite{MemJam} uses read-after-write false dependencies to introduce latency when a victim program reads data with a specific page offset.
By measuring the run time of the victim program a high number of times while \emph{jamming} different page offsets, the attacker can infer which offsets are read more often by the victim.
This attack can leak information with a four byte granularity, but requires an extremely high number of runs (\emph{50~million runs} for an attack against a simple and deterministic SGX enclave).
However, with \abbrev, the page offsets of data change between different runs, making the correlation of timing information for different runs exponentially more involved.
Moreover, the accesses due to \abbrev's own code generate a significant amount of noise, which complicates the matter further.
Finally, the code of \abbrev itself was designed to not be vulnerable to MemJam attacks, e.g., by randomizing the permutation buffer layout (see~\Cref{sec:permubuffer}).

\paragraph{Leakage quantification.} Quantification of cache-based information leakage has been studied in previous works. For example, CacheAudit~\cite{doychev2015cacheaudit} is a well-known static analysis framework that given an x86 binary and a cache configuration yields an upper bound on the amount of information leakage via cache- and time-based side-channels. The information leakage is quantified based on the number of side-channel observations an attacker can obtain. 

CacheAudit, and similar existing tools, are not applicable to our scenario for two main reasons. First, 
in the model of CacheAudit, randomly permuted observations contribute to the total number of observations, even though the attacker may not learn any useful information from such accesses. Second, CacheAudit does not consider information leakage through other channels, such as page faults, that can be correlated with cache observations. Therefore, CacheAudit cannot be used to quantify informations leakage of \abbrev.

\section{Conclusion} 
\label{sec:conclusion}

In this paper we have proposed semantic-agnostic data randomization as a new defensive approach against side-channel attacks on SGX. We have designed and implemented \abbrev, which allows to instrument enclave code such that all data locations in enclave memory are permuted at cache-line granularity and re-randomized at runtime. Unlike previous defenses, our solution allows non-expert developers to harden their enclaves against various data-driven attack strategies with an adjustable security-performance trade-off.

\section*{Acknowledgments}
The authors would like to thank Urs M\"uller for his feedback in the initial discussions that led to this work.

\noindent This work has been supported by the German Research Foundation (DFG) as part of projects HWSec, P3 and S2 within the CRC 1119 CROSSING, by the German Federal Ministry of Education and Research (BMBF) and the Hessen State Ministry for Higher Education, Research and the Arts (HMWK) within CRISP, by BMBF within the projects iBlockchain and CloudProtect, and by the Intel Collaborative Research Institute for Collaborative Autonomous \& Resilient Systems (ICRI-CARS).

\bibliographystyle{ACM-Reference-Format}
\bibliography{bibliography}


\begin{thebibliography}{78}


\ifx \showCODEN    \undefined \def \showCODEN     #1{\unskip}     \fi
\ifx \showDOI      \undefined \def \showDOI       #1{#1}\fi
\ifx \showISBNx    \undefined \def \showISBNx     #1{\unskip}     \fi
\ifx \showISBNxiii \undefined \def \showISBNxiii  #1{\unskip}     \fi
\ifx \showISSN     \undefined \def \showISSN      #1{\unskip}     \fi
\ifx \showLCCN     \undefined \def \showLCCN      #1{\unskip}     \fi
\ifx \shownote     \undefined \def \shownote      #1{#1}          \fi
\ifx \showarticletitle \undefined \def \showarticletitle #1{#1}   \fi
\ifx \showURL      \undefined \def \showURL       {\relax}        \fi
\providecommand\bibfield[2]{#2}
\providecommand\bibinfo[2]{#2}
\providecommand\natexlab[1]{#1}
\providecommand\showeprint[2][]{arXiv:#2}

\bibitem[\protect\citeauthoryear{Ahmad, Joe, Xiao, Zhang, Shin, and Lee}{Ahmad
  et~al\mbox{.}}{2019}]%
        {obfuscuro}
\bibfield{author}{\bibinfo{person}{Adil Ahmad}, \bibinfo{person}{Byunggill
  Joe}, \bibinfo{person}{Yuan Xiao}, \bibinfo{person}{Yinqian Zhang},
  \bibinfo{person}{Insik Shin}, {and} \bibinfo{person}{Byoungyoung Lee}.}
  \bibinfo{year}{2019}\natexlab{}.
\newblock \showarticletitle{{Obfuscuro: A Commodity Obfuscation Engine on Intel
  SGX}}. In \bibinfo{booktitle}{\emph{Network and Distributed System Security
  Symposium}}.
\newblock


\bibitem[\protect\citeauthoryear{{ARM Limited}}{{ARM Limited}}{2009}]%
        {trustzone}
\bibfield{author}{\bibinfo{person}{{ARM Limited}}.}
  \bibinfo{year}{2009}\natexlab{}.
\newblock \bibinfo{title}{{ARM Security Technology -- Building a Secure System
  using TrustZone Technology}}.
\newblock
  \bibinfo{howpublished}{\url{http://infocenter.arm.com/help/topic/com.arm.doc.prd29-genc-009492c/PRD29-GENC-009492C_trustzone_security_whitepaper.pdf}}.
\newblock


\bibitem[\protect\citeauthoryear{Batina, Jauernig, Mentens, Sadeghi, and
  Stapf}{Batina et~al\mbox{.}}{2019}]%
        {batina2019hardware}
\bibfield{author}{\bibinfo{person}{Lejla Batina}, \bibinfo{person}{Patrick
  Jauernig}, \bibinfo{person}{Nele Mentens}, \bibinfo{person}{A-R Sadeghi},
  {and} \bibinfo{person}{Emmanuel Stapf}.} \bibinfo{year}{2019}\natexlab{}.
\newblock \showarticletitle{In Hardware We Trust: Gains and Pains of
  Hardware-assisted Security}.
\newblock  (\bibinfo{year}{2019}).
\newblock


\bibitem[\protect\citeauthoryear{Baumann, Peinado, and Hunt}{Baumann
  et~al\mbox{.}}{2014}]%
        {haven}
\bibfield{author}{\bibinfo{person}{Andrew Baumann}, \bibinfo{person}{Marcus
  Peinado}, {and} \bibinfo{person}{Galen Hunt}.}
  \bibinfo{year}{2014}\natexlab{}.
\newblock \showarticletitle{{Shielding Applications from an Untrusted Cloud
  with Haven}}.
\newblock


\bibitem[\protect\citeauthoryear{Bellare, Rogaway, and Spies}{Bellare
  et~al\mbox{.}}{2010}]%
        {ffx}
\bibfield{author}{\bibinfo{person}{Mihir Bellare}, \bibinfo{person}{Phillip
  Rogaway}, {and} \bibinfo{person}{Terence Spies}.}
  \bibinfo{year}{2010}\natexlab{}.
\newblock \bibinfo{booktitle}{\emph{{The FFX Mode of Operation for
  Format-Preserving Encryption}}}.
\newblock \bibinfo{type}{{T}echnical {R}eport}.
\newblock


\bibitem[\protect\citeauthoryear{Bigelow, Hobson, Rudd, Streilein, and
  Okhravi}{Bigelow et~al\mbox{.}}{2015}]%
        {bigelow2015timely}
\bibfield{author}{\bibinfo{person}{David Bigelow}, \bibinfo{person}{Thomas
  Hobson}, \bibinfo{person}{Robert Rudd}, \bibinfo{person}{William Streilein},
  {and} \bibinfo{person}{Hamed Okhravi}.} \bibinfo{year}{2015}\natexlab{}.
\newblock \showarticletitle{Timely rerandomization for mitigating memory
  disclosures}. In \bibinfo{booktitle}{\emph{ACM SIGSAC Conference on Computer
  and Communications Security}}. ACM.
\newblock


\bibitem[\protect\citeauthoryear{Biondo, Conti, Davi, Frassetto, and
  Sadeghi}{Biondo et~al\mbox{.}}{2018}]%
        {guards-dilemma}
\bibfield{author}{\bibinfo{person}{Andrea Biondo}, \bibinfo{person}{Mauro
  Conti}, \bibinfo{person}{Lucas Davi}, \bibinfo{person}{Tommaso Frassetto},
  {and} \bibinfo{person}{Ahmad-Reza Sadeghi}.} \bibinfo{year}{2018}\natexlab{}.
\newblock \showarticletitle{The Guard's Dilemma: Efficient Code-Reuse Attacks
  Against Intel SGX}. In \bibinfo{booktitle}{\emph{27th USENIX Security
  Symposium}}.
\newblock


\bibitem[\protect\citeauthoryear{Brasser, Davi, Dhavlle, Frassetto, Dinakarrao,
  Rafatirad, Sadeghi, Sasan, Sayadi, Zeitouni, et~al\mbox{.}}{Brasser
  et~al\mbox{.}}{2018a}]%
        {brasser2018advances}
\bibfield{author}{\bibinfo{person}{Ferdinand Brasser}, \bibinfo{person}{Lucas
  Davi}, \bibinfo{person}{Abhijitt Dhavlle}, \bibinfo{person}{Tommaso
  Frassetto}, \bibinfo{person}{Sai Manoj~Pudukotai Dinakarrao},
  \bibinfo{person}{Setareh Rafatirad}, \bibinfo{person}{Ahmad-Reza Sadeghi},
  \bibinfo{person}{Avesta Sasan}, \bibinfo{person}{Hossein Sayadi},
  \bibinfo{person}{Shaza Zeitouni}, {et~al\mbox{.}}}
  \bibinfo{year}{2018}\natexlab{a}.
\newblock \showarticletitle{Advances and throwbacks in hardware-assisted
  security: special session}. In \bibinfo{booktitle}{\emph{Proceedings of the
  International Conference on Compilers, Architecture and Synthesis for
  Embedded Systems}}. IEEE Press, \bibinfo{pages}{15}.
\newblock


\bibitem[\protect\citeauthoryear{Brasser, Frassetto, Riedhammer, Sadeghi,
  Schneider, and Weinert}{Brasser et~al\mbox{.}}{2018b}]%
        {voiceguard}
\bibfield{author}{\bibinfo{person}{Ferdinand Brasser}, \bibinfo{person}{Tommaso
  Frassetto}, \bibinfo{person}{Korbinian Riedhammer},
  \bibinfo{person}{Ahmad-Reza Sadeghi}, \bibinfo{person}{Thomas Schneider},
  {and} \bibinfo{person}{Christian Weinert}.} \bibinfo{year}{2018}\natexlab{b}.
\newblock \showarticletitle{VoiceGuard: Secure and Private Speech Processing}.
  In \bibinfo{booktitle}{\emph{Interspeech 2018}}.
  \bibinfo{publisher}{International Speech Communication Association (ISCA)},
  \bibinfo{pages}{1303--1307}.
\newblock


\bibitem[\protect\citeauthoryear{Brasser, Gens, Jauernig, Sadeghi, and
  Stapf}{Brasser et~al\mbox{.}}{2019}]%
        {sanctuary}
\bibfield{author}{\bibinfo{person}{Ferdinand Brasser}, \bibinfo{person}{David
  Gens}, \bibinfo{person}{Patrick Jauernig}, \bibinfo{person}{Ahmad-Reza
  Sadeghi}, {and} \bibinfo{person}{Emmanuel Stapf}.}
  \bibinfo{year}{2019}\natexlab{}.
\newblock \showarticletitle{SANCTUARY: ARMing TrustZone with User-space
  Enclaves}. In \bibinfo{booktitle}{\emph{26th Annual Network \& Distributed
  System Security Symposium (NDSS)}}.
\newblock


\bibitem[\protect\citeauthoryear{Brasser, M{\"u}ller, Dmitrienko, Kostiainen,
  Capkun, and Sadeghi}{Brasser et~al\mbox{.}}{2017}]%
        {BMD+17b}
\bibfield{author}{\bibinfo{person}{Ferdinand Brasser}, \bibinfo{person}{Urs
  M{\"u}ller}, \bibinfo{person}{Alexandra Dmitrienko}, \bibinfo{person}{Kari
  Kostiainen}, \bibinfo{person}{Srdjan Capkun}, {and}
  \bibinfo{person}{Ahmad-Reza Sadeghi}.} \bibinfo{year}{2017}\natexlab{}.
\newblock \showarticletitle{Software Grand Exposure: {SGX} Cache Attacks Are
  Practical}. In \bibinfo{booktitle}{\emph{{USENIX} Workshop on Offensive
  Technologies}}.
\newblock


\bibitem[\protect\citeauthoryear{Brickell, Graunke, and Seifert}{Brickell
  et~al\mbox{.}}{2006}]%
        {BrGrSe2006}
\bibfield{author}{\bibinfo{person}{E. Brickell}, \bibinfo{person}{G. Graunke},
  {and} \bibinfo{person}{J.-P. Seifert}.} \bibinfo{year}{2006}\natexlab{}.
\newblock \showarticletitle{Mitigating cache/timing attacks in {AES} and {RSA}
  software implementations}. In \bibinfo{booktitle}{\emph{RSA Conference 2006,
  session DEV-203}}.
\newblock


\bibitem[\protect\citeauthoryear{{BYTE~Magazine} and Mayer}{{BYTE~Magazine} and
  Mayer}{2011}]%
        {nbench}
\bibfield{author}{\bibinfo{person}{{BYTE~Magazine}} {and}
  \bibinfo{person}{Uwe~F. Mayer}.} \bibinfo{year}{1995-2011}\natexlab{}.
\newblock \bibinfo{title}{{BYTEmark} benchmark (nbench), port to Linux}.
\newblock
\newblock
\newblock
\shownote{Original address \url{http://www.tux.org/~mayer/linux/bmark.html},
  now archived at
  \url{https://web.archive.org/web/20151215162836/http://www.tux.org/~mayer/linux/bmark.html}.}


\bibitem[\protect\citeauthoryear{Catuogno, Dmitrienko, Eriksson, Kuhlmann,
  Ramunno, Sadeghi, Schulz, Schunter, Winandy, and Zhan}{Catuogno
  et~al\mbox{.}}{2009}]%
        {CDE+09}
\bibfield{author}{\bibinfo{person}{Luigi Catuogno}, \bibinfo{person}{Alexandra
  Dmitrienko}, \bibinfo{person}{Konrad Eriksson}, \bibinfo{person}{Dirk
  Kuhlmann}, \bibinfo{person}{Gianluca Ramunno}, \bibinfo{person}{Ahmad-Reza
  Sadeghi}, \bibinfo{person}{Steffen Schulz}, \bibinfo{person}{Matthias
  Schunter}, \bibinfo{person}{Marcel Winandy}, {and} \bibinfo{person}{Jing
  Zhan}.} \bibinfo{year}{2009}\natexlab{}.
\newblock \showarticletitle{{Trusted Virtual Domains -- Design, Implementation
  and Lessons Learned}}. In \bibinfo{booktitle}{\emph{International Conference
  on Trusted Systems}}.
\newblock


\bibitem[\protect\citeauthoryear{Chandra, Karande, Lin, Khan, Kantarcioglu, and
  Thuraisingham}{Chandra et~al\mbox{.}}{2017}]%
        {CKL+17}
\bibfield{author}{\bibinfo{person}{Swarup Chandra}, \bibinfo{person}{Vishal
  Karande}, \bibinfo{person}{Zhiqiang Lin}, \bibinfo{person}{Latifur Khan},
  \bibinfo{person}{Murat Kantarcioglu}, {and} \bibinfo{person}{Bhavani
  Thuraisingham}.} \bibinfo{year}{2017}\natexlab{}.
\newblock \showarticletitle{{Securing Data Analytics on SGX with
  Randomization}}. In \bibinfo{booktitle}{\emph{European Symposium on Research
  in Computer Security}}.
\newblock


\bibitem[\protect\citeauthoryear{Chen, Chen, Xiao, Zhang, Lin, and Lai}{Chen
  et~al\mbox{.}}{2018}]%
        {sgxpectre}
\bibfield{author}{\bibinfo{person}{Guoxing Chen}, \bibinfo{person}{Sanchuan
  Chen}, \bibinfo{person}{Yuan Xiao}, \bibinfo{person}{Yinqian Zhang},
  \bibinfo{person}{Zhiqiang Lin}, {and} \bibinfo{person}{Ten~H. Lai}.}
  \bibinfo{year}{2018}\natexlab{}.
\newblock \bibinfo{title}{{SgxPectre Attacks: Stealing Intel Secrets from SGX
  Enclaves via Speculative Execution}}.
\newblock
\newblock
\showeprint{arXiv:1802.09085v3}


\bibitem[\protect\citeauthoryear{Chen, Zhang, Reiter, and Zhang}{Chen
  et~al\mbox{.}}{2017}]%
        {incognito2017}
\bibfield{author}{\bibinfo{person}{Sanchuan Chen}, \bibinfo{person}{Xiaokuan
  Zhang}, \bibinfo{person}{Michael~K. Reiter}, {and} \bibinfo{person}{Yinqian
  Zhang}.} \bibinfo{year}{2017}\natexlab{}.
\newblock \showarticletitle{Detecting Privileged Side-Channel Attacks in
  Shielded Execution with {D{\'e}j{\'a} Vu}}. In \bibinfo{booktitle}{\emph{ACM
  Symposium on Information, Computer and Communications Security}}.
\newblock


\bibitem[\protect\citeauthoryear{Costan and Devadas}{Costan and
  Devadas}{2016}]%
        {Cos2016}
\bibfield{author}{\bibinfo{person}{Victor Costan} {and}
  \bibinfo{person}{Srinivas Devadas}.} \bibinfo{year}{2016}\natexlab{}.
\newblock \bibinfo{booktitle}{\emph{{Intel SGX} Explained}}.
\newblock \bibinfo{type}{{T}echnical {R}eport}.
  \bibinfo{institution}{Cryptology ePrint Archive. Report 2016/086}.
\newblock
\newblock
\shownote{\url{https://eprint.iacr.org/2016/086.pdf}.}


\bibitem[\protect\citeauthoryear{Crane, Homescu, Brunthaler, Larsen, and
  Franz}{Crane et~al\mbox{.}}{2015}]%
        {CHB+15}
\bibfield{author}{\bibinfo{person}{Stephen Crane}, \bibinfo{person}{Andrei
  Homescu}, \bibinfo{person}{Stefan Brunthaler}, \bibinfo{person}{Per Larsen},
  {and} \bibinfo{person}{Michael Franz}.} \bibinfo{year}{2015}\natexlab{}.
\newblock \showarticletitle{Thwarting Cache Side-Channel Attacks Through
  Dynamic Software Diversity}. In \bibinfo{booktitle}{\emph{Network and
  Distributed System Security Symposium}}.
\newblock


\bibitem[\protect\citeauthoryear{Das, Eckey, Frassetto, Gens,
  Host{\'a}kov{\'a}, Jauernig, Faust, and Sadeghi}{Das et~al\mbox{.}}{2019}]%
        {fastkitten}
\bibfield{author}{\bibinfo{person}{Poulami Das}, \bibinfo{person}{Lisa Eckey},
  \bibinfo{person}{Tommaso Frassetto}, \bibinfo{person}{David Gens},
  \bibinfo{person}{Kristina Host{\'a}kov{\'a}}, \bibinfo{person}{Patrick
  Jauernig}, \bibinfo{person}{Sebastian Faust}, {and}
  \bibinfo{person}{Ahmad-Reza Sadeghi}.} \bibinfo{year}{2019}\natexlab{}.
\newblock \showarticletitle{FastKitten: Practical Smart Contracts on Bitcoin}.
  In \bibinfo{booktitle}{\emph{28th USENIX Security Symposium}}.
\newblock


\bibitem[\protect\citeauthoryear{Davi, Dmitrienko, N{\"u}rnberger, and
  Sadeghi}{Davi et~al\mbox{.}}{2013}]%
        {TUD-CS-2013-0042}
\bibfield{author}{\bibinfo{person}{Lucas Davi}, \bibinfo{person}{Alexandra
  Dmitrienko}, \bibinfo{person}{Stefan N{\"u}rnberger}, {and}
  \bibinfo{person}{Ahmad-Reza Sadeghi}.} \bibinfo{year}{2013}\natexlab{}.
\newblock \showarticletitle{Gadge Me If You Can - Secure and Efficient Ad-hoc
  Instruction-Level Randomization for x86 and {ARM}}. In
  \bibinfo{booktitle}{\emph{ACM Symposium on Information, Computer and
  Communications Security}}.
\newblock


\bibitem[\protect\citeauthoryear{Dessouky, Frassetto, and Sadeghi}{Dessouky
  et~al\mbox{.}}{2020}]%
        {hybcache}
\bibfield{author}{\bibinfo{person}{Ghada Dessouky}, \bibinfo{person}{Tommaso
  Frassetto}, {and} \bibinfo{person}{Ahmad-Reza Sadeghi}.}
  \bibinfo{year}{2020}\natexlab{}.
\newblock \showarticletitle{HybCache: Hybrid Side-Channel-Resilient Caches for
  Trusted Execution Environments}. In \bibinfo{booktitle}{\emph{29th {USENIX}
  Security Symposium}}.
\newblock


\bibitem[\protect\citeauthoryear{Domnitser, Jaleel, Loew, Abu-Ghazaleh, and
  Ponomarev}{Domnitser et~al\mbox{.}}{2012a}]%
        {domnitser2012non}
\bibfield{author}{\bibinfo{person}{Leonid Domnitser}, \bibinfo{person}{Aamer
  Jaleel}, \bibinfo{person}{Jason Loew}, \bibinfo{person}{Nael Abu-Ghazaleh},
  {and} \bibinfo{person}{Dmitry Ponomarev}.} \bibinfo{year}{2012}\natexlab{a}.
\newblock \showarticletitle{Non-monopolizable caches: Low-complexity mitigation
  of cache side channel attacks}.
\newblock \bibinfo{journal}{\emph{ACM Transactions on Architecture and Code
  Optimization (TACO)}} \bibinfo{volume}{8}, \bibinfo{number}{4}
  (\bibinfo{year}{2012}).
\newblock


\bibitem[\protect\citeauthoryear{Domnitser, Jaleel, Loew, Abu-Ghazaleh, and
  Ponomarev}{Domnitser et~al\mbox{.}}{2012b}]%
        {DoJaLoAbPo2012}
\bibfield{author}{\bibinfo{person}{Leonid Domnitser}, \bibinfo{person}{Aamer
  Jaleel}, \bibinfo{person}{Jason Loew}, \bibinfo{person}{Nael Abu-Ghazaleh},
  {and} \bibinfo{person}{Dmitry Ponomarev}.} \bibinfo{year}{2012}\natexlab{b}.
\newblock \showarticletitle{Non-monopolizable Caches: Low-complexity Mitigation
  of Cache Side Channel Attacks}.
\newblock \bibinfo{journal}{\emph{ACM Transactions on Architecture and Code
  Optimization}} (\bibinfo{year}{2012}).
\newblock
\urldef\tempurl%
\url{https://doi.org/10.1145/2086696.2086714}
\showDOI{\tempurl}


\bibitem[\protect\citeauthoryear{Doychev, K{\"o}pf, Mauborgne, and
  Reineke}{Doychev et~al\mbox{.}}{2015}]%
        {doychev2015cacheaudit}
\bibfield{author}{\bibinfo{person}{Goran Doychev}, \bibinfo{person}{Boris
  K{\"o}pf}, \bibinfo{person}{Laurent Mauborgne}, {and} \bibinfo{person}{Jan
  Reineke}.} \bibinfo{year}{2015}\natexlab{}.
\newblock \showarticletitle{Cacheaudit: A tool for the static analysis of cache
  side channels}.
\newblock \bibinfo{journal}{\emph{ACM Transactions on Information and System
  Security (TISSEC)}} \bibinfo{volume}{18}, \bibinfo{number}{1}
  (\bibinfo{year}{2015}).
\newblock


\bibitem[\protect\citeauthoryear{Frassetto, Gens, Liebchen, and
  Sadeghi}{Frassetto et~al\mbox{.}}{2017}]%
        {jitguard}
\bibfield{author}{\bibinfo{person}{Tommaso Frassetto}, \bibinfo{person}{David
  Gens}, \bibinfo{person}{Christopher Liebchen}, {and}
  \bibinfo{person}{Ahmad-Reza Sadeghi}.} \bibinfo{year}{2017}\natexlab{}.
\newblock \showarticletitle{JITGuard: Hardening Just-in-time Compilers with
  SGX}. In \bibinfo{booktitle}{\emph{24th ACM Conference on Computer and
  Communications Security (CCS)}}.
\newblock
\showISBNx{978-1-4503-4946-8/17/10}


\bibitem[\protect\citeauthoryear{Fuhry, Bahmani, Brasser, Hahn, Kerschbaum, and
  Sadeghi}{Fuhry et~al\mbox{.}}{2017}]%
        {hardidx}
\bibfield{author}{\bibinfo{person}{Benny Fuhry}, \bibinfo{person}{Raad
  Bahmani}, \bibinfo{person}{Ferdinand Brasser}, \bibinfo{person}{Florian
  Hahn}, \bibinfo{person}{Florian Kerschbaum}, {and}
  \bibinfo{person}{Ahmad-Reza Sadeghi}.} \bibinfo{year}{2017}\natexlab{}.
\newblock \showarticletitle{HardIDX: Practical and Secure Index with SGX}. In
  \bibinfo{booktitle}{\emph{Conference on Data and Applications Security and
  Privacy (DBSec)}}.
\newblock


\bibitem[\protect\citeauthoryear{Goldreich}{Goldreich}{1987}]%
        {goldreich1987towards}
\bibfield{author}{\bibinfo{person}{Oded Goldreich}.}
  \bibinfo{year}{1987}\natexlab{}.
\newblock \showarticletitle{Towards a theory of software protection and
  simulation by oblivious {RAMs}}. In \bibinfo{booktitle}{\emph{Annual ACM
  Symposium on Theory of Computing}}. ACM.
\newblock


\bibitem[\protect\citeauthoryear{Goldreich and Ostrovsky}{Goldreich and
  Ostrovsky}{1996}]%
        {GolOst1996}
\bibfield{author}{\bibinfo{person}{Oded Goldreich} {and}
  \bibinfo{person}{Rafail Ostrovsky}.} \bibinfo{year}{1996}\natexlab{}.
\newblock \showarticletitle{Software Protection and Simulation on Oblivious
  {RAMs}}.
\newblock \bibinfo{journal}{\emph{J. ACM}} (\bibinfo{year}{1996}).
\newblock


\bibitem[\protect\citeauthoryear{Goodrich, Mitzenmacher, Ohrimenko, and
  Tamassia}{Goodrich et~al\mbox{.}}{2012}]%
        {goodrich2012privacy}
\bibfield{author}{\bibinfo{person}{Michael~T Goodrich},
  \bibinfo{person}{Michael Mitzenmacher}, \bibinfo{person}{Olga Ohrimenko},
  {and} \bibinfo{person}{Roberto Tamassia}.} \bibinfo{year}{2012}\natexlab{}.
\newblock \showarticletitle{Privacy-preserving group data access via stateless
  oblivious {RAM} simulation}. In \bibinfo{booktitle}{\emph{Annual ACM-SIAM
  symposium on Discrete Algorithms}}. Society for Industrial and Applied
  Mathematics.
\newblock


\bibitem[\protect\citeauthoryear{G\"{o}tzfried, Eckert, Schinzel, and
  M\"{u}ller}{G\"{o}tzfried et~al\mbox{.}}{2017}]%
        {GES+17}
\bibfield{author}{\bibinfo{person}{Johannes G\"{o}tzfried},
  \bibinfo{person}{Moritz Eckert}, \bibinfo{person}{Sebastian Schinzel}, {and}
  \bibinfo{person}{Tilo M\"{u}ller}.} \bibinfo{year}{2017}\natexlab{}.
\newblock \showarticletitle{{Cache Attacks on Intel SGX}}. In
  \bibinfo{booktitle}{\emph{European Workshop on Systems Security}}.
\newblock


\bibitem[\protect\citeauthoryear{Gras, Razavi, Bos, and Giuffrida}{Gras
  et~al\mbox{.}}{2018}]%
        {GRB+18}
\bibfield{author}{\bibinfo{person}{Ben Gras}, \bibinfo{person}{Kaveh Razavi},
  \bibinfo{person}{Herbert Bos}, {and} \bibinfo{person}{Cristiano Giuffrida}.}
  \bibinfo{year}{2018}\natexlab{}.
\newblock \showarticletitle{{Translation Leak-aside Buffer: Defeating Cache
  Side-channel Protections with {TLB} Attacks}}. In
  \bibinfo{booktitle}{\emph{27th {USENIX} Security Symposium}}.
\newblock
\showISBNx{978-1-931971-46-1}
\urldef\tempurl%
\url{https://www.usenix.org/conference/usenixsecurity18/presentation/gras}
\showURL{%
\tempurl}


\bibitem[\protect\citeauthoryear{Gruss, Lettner, Schuster, Ohrimenko, Haller,
  and Costa}{Gruss et~al\mbox{.}}{2017}]%
        {cloak}
\bibfield{author}{\bibinfo{person}{Daniel Gruss}, \bibinfo{person}{Julian
  Lettner}, \bibinfo{person}{Felix Schuster}, \bibinfo{person}{Olya Ohrimenko},
  \bibinfo{person}{Istvan Haller}, {and} \bibinfo{person}{Manuel Costa}.}
  \bibinfo{year}{2017}\natexlab{}.
\newblock \showarticletitle{{Strong and Efficient Cache Side-Channel Protection
  using Hardware Transactional Memory}}. In \bibinfo{booktitle}{\emph{26th
  {USENIX} Security Symposium}}.
\newblock


\bibitem[\protect\citeauthoryear{Hiser, Nguyen-Tuong, Co, Hall, and
  Davidson}{Hiser et~al\mbox{.}}{2012}]%
        {Hi2012}
\bibfield{author}{\bibinfo{person}{Jason~D. Hiser}, \bibinfo{person}{Anh
  Nguyen-Tuong}, \bibinfo{person}{Michele Co}, \bibinfo{person}{Matthew Hall},
  {and} \bibinfo{person}{Jack~W. Davidson}.} \bibinfo{year}{2012}\natexlab{}.
\newblock \showarticletitle{{ILR}: Where'd My Gadgets Go?}. In
  \bibinfo{booktitle}{\emph{IEEE Symposium on Security and Privacy}}.
\newblock


\bibitem[\protect\citeauthoryear{{Intel}}{{Intel}}{2015}]%
        {ISCA2015}
\bibfield{author}{\bibinfo{person}{{Intel}}.} \bibinfo{year}{2015}\natexlab{}.
\newblock \bibinfo{title}{Intel {Software Guard Extensions.} {Tutorial}
  slides}.
\newblock
  \bibinfo{howpublished}{\url{https://software.intel.com/sites/default/files/332680-002.pdf}}.
\newblock
\newblock
\shownote{Reference Number: 332680-002, revision 1.1.}


\bibitem[\protect\citeauthoryear{{Intel}}{{Intel}}{2016}]%
        {Intel-manual}
\bibfield{author}{\bibinfo{person}{{Intel}}.} \bibinfo{year}{2016}\natexlab{}.
\newblock \bibinfo{title}{{Intel 64 and IA-32} Architectures Software
  Developer’s Manual}.
\newblock
  \bibinfo{howpublished}{\url{http://www.intel.com/content/www/us/en/architecture-and-technology/64-ia-32-architectures-software-developer-manual-325462.html}}.
\newblock


\bibitem[\protect\citeauthoryear{Keramidas, Antonopoulos, Serpanos, and
  Kaxiras}{Keramidas et~al\mbox{.}}{2008}]%
        {Keramidas2008}
\bibfield{author}{\bibinfo{person}{G. Keramidas}, \bibinfo{person}{A.
  Antonopoulos}, \bibinfo{person}{D.~N. Serpanos}, {and} \bibinfo{person}{S.
  Kaxiras}.} \bibinfo{year}{2008}\natexlab{}.
\newblock \showarticletitle{Non deterministic caches: {A} simple and effective
  defense against side channel attacks}.
\newblock \bibinfo{journal}{\emph{Design Automation for Embedded Systems}}
  (\bibinfo{year}{2008}).
\newblock


\bibitem[\protect\citeauthoryear{Kil, Jun, Bookholt, Xu, and Ning}{Kil
  et~al\mbox{.}}{2006}]%
        {kil2006address}
\bibfield{author}{\bibinfo{person}{Chongkyung Kil}, \bibinfo{person}{Jinsuk
  Jun}, \bibinfo{person}{Christopher Bookholt}, \bibinfo{person}{Jun Xu}, {and}
  \bibinfo{person}{Peng Ning}.} \bibinfo{year}{2006}\natexlab{}.
\newblock \showarticletitle{Address Space Layout Permutation {(ASLP)}: Towards
  Fine-Grained Randomization of Commodity Software}. In
  \bibinfo{booktitle}{\emph{Annual Computer Security Applications Conference}}.
\newblock


\bibitem[\protect\citeauthoryear{Kocher, Horn, Fogh, , Genkin, Gruss, Haas,
  Hamburg, Lipp, Mangard, Prescher, Schwarz, and Yarom}{Kocher
  et~al\mbox{.}}{2019}]%
        {Kocher2018spectre}
\bibfield{author}{\bibinfo{person}{Paul Kocher}, \bibinfo{person}{Jann Horn},
  \bibinfo{person}{Anders Fogh}, \bibinfo{person}{}, \bibinfo{person}{Daniel
  Genkin}, \bibinfo{person}{Daniel Gruss}, \bibinfo{person}{Werner Haas},
  \bibinfo{person}{Mike Hamburg}, \bibinfo{person}{Moritz Lipp},
  \bibinfo{person}{Stefan Mangard}, \bibinfo{person}{Thomas Prescher},
  \bibinfo{person}{Michael Schwarz}, {and} \bibinfo{person}{Yuval Yarom}.}
  \bibinfo{year}{2019}\natexlab{}.
\newblock \showarticletitle{{Spectre Attacks: Exploiting Speculative
  Execution}}. In \bibinfo{booktitle}{\emph{40th IEEE Symposium on Security and
  Privacy (S\&P'19)}}.
\newblock


\bibitem[\protect\citeauthoryear{Kong, Acii{\c{c}}mez, Seifert, and Zhou}{Kong
  et~al\mbox{.}}{2009}]%
        {kong2009hardware}
\bibfield{author}{\bibinfo{person}{Jingfei Kong}, \bibinfo{person}{Onur
  Acii{\c{c}}mez}, \bibinfo{person}{Jean-Pierre Seifert}, {and}
  \bibinfo{person}{Huiyang Zhou}.} \bibinfo{year}{2009}\natexlab{}.
\newblock \showarticletitle{Hardware-software integrated approaches to defend
  against software cache-based side channel attacks}. In
  \bibinfo{booktitle}{\emph{IEEE International Symposium on High Performance
  Computer Architecture}}. IEEE.
\newblock


\bibitem[\protect\citeauthoryear{K\"{u}\c{c}\"{u}k, Paverd, Martin, Asokan,
  Simpson, and Ankele}{K\"{u}\c{c}\"{u}k et~al\mbox{.}}{2016}]%
        {KPM+16}
\bibfield{author}{\bibinfo{person}{Kubilay~Ahmet K\"{u}\c{c}\"{u}k},
  \bibinfo{person}{Andrew Paverd}, \bibinfo{person}{Andrew Martin},
  \bibinfo{person}{N. Asokan}, \bibinfo{person}{Andrew Simpson}, {and}
  \bibinfo{person}{Robin Ankele}.} \bibinfo{year}{2016}\natexlab{}.
\newblock \showarticletitle{{Exploring the Use of Intel SGX for Secure
  Many-Party Applications}}.
\newblock


\bibitem[\protect\citeauthoryear{Lee, Shih, Gera, Kim, Kim, and Peinado}{Lee
  et~al\mbox{.}}{2017}]%
        {LSG+17}
\bibfield{author}{\bibinfo{person}{Sangho Lee}, \bibinfo{person}{Ming-Wei
  Shih}, \bibinfo{person}{Prasun Gera}, \bibinfo{person}{Taesoo Kim},
  \bibinfo{person}{Hyesoon Kim}, {and} \bibinfo{person}{Marcus Peinado}.}
  \bibinfo{year}{2017}\natexlab{}.
\newblock \showarticletitle{{Inferring Fine-grained Control Flow Inside {SGX}
  Enclaves with Branch Shadowing}}. In \bibinfo{booktitle}{\emph{USENIX
  Security Symposium}}.
\newblock


\bibitem[\protect\citeauthoryear{Lipp, Schwarz, Gruss, Prescher, Haas, Fogh,
  Horn, Mangard, Kocher, Genkin, Yarom, and Hamburg}{Lipp
  et~al\mbox{.}}{2018}]%
        {Lipp2018meltdown}
\bibfield{author}{\bibinfo{person}{Moritz Lipp}, \bibinfo{person}{Michael
  Schwarz}, \bibinfo{person}{Daniel Gruss}, \bibinfo{person}{Thomas Prescher},
  \bibinfo{person}{Werner Haas}, \bibinfo{person}{Anders Fogh},
  \bibinfo{person}{Jann Horn}, \bibinfo{person}{Stefan Mangard},
  \bibinfo{person}{Paul Kocher}, \bibinfo{person}{Daniel Genkin},
  \bibinfo{person}{Yuval Yarom}, {and} \bibinfo{person}{Mike Hamburg}.}
  \bibinfo{year}{2018}\natexlab{}.
\newblock \showarticletitle{{Meltdown: Reading Kernel Memory from User Space}}.
  In \bibinfo{booktitle}{\emph{27th {USENIX} Security Symposium ({USENIX}
  Security 18)}}.
\newblock


\bibitem[\protect\citeauthoryear{Liu, Harris, Maas, Hicks, Tiwari, and Shi}{Liu
  et~al\mbox{.}}{2015}]%
        {liu2015ghostrider}
\bibfield{author}{\bibinfo{person}{Chang Liu}, \bibinfo{person}{Austin Harris},
  \bibinfo{person}{Martin Maas}, \bibinfo{person}{Michael Hicks},
  \bibinfo{person}{Mohit Tiwari}, {and} \bibinfo{person}{Elaine Shi}.}
  \bibinfo{year}{2015}\natexlab{}.
\newblock \showarticletitle{{Ghostrider: A hardware-software system for memory
  trace oblivious computation}}.
\newblock \bibinfo{journal}{\emph{ACM SIGARCH Computer Architecture News}}
  \bibinfo{volume}{43}, \bibinfo{number}{1} (\bibinfo{year}{2015}).
\newblock


\bibitem[\protect\citeauthoryear{Liu, Hicks, and Shi}{Liu
  et~al\mbox{.}}{2013}]%
        {liu-csf13}
\bibfield{author}{\bibinfo{person}{Chang Liu}, \bibinfo{person}{Michael Hicks},
  {and} \bibinfo{person}{Elaine Shi}.} \bibinfo{year}{2013}\natexlab{}.
\newblock \showarticletitle{Memory trace oblivious program execution}. In
  \bibinfo{booktitle}{\emph{IEEE Computer Security Foundations Symposium}}.
\newblock


\bibitem[\protect\citeauthoryear{Liu and Lee}{Liu and Lee}{2014}]%
        {LiLe2014}
\bibfield{author}{\bibinfo{person}{Fangfei Liu} {and} \bibinfo{person}{Ruby~B.
  Lee}.} \bibinfo{year}{2014}\natexlab{}.
\newblock \showarticletitle{Random Fill Cache Architecture}. In
  \bibinfo{booktitle}{\emph{47th Annual IEEE/ACM International Symposium on
  Microarchitecture}}.
\newblock


\bibitem[\protect\citeauthoryear{{LLVM Foundation}}{{LLVM Foundation}}{2019}]%
        {llvm}
\bibfield{author}{\bibinfo{person}{{LLVM Foundation}}.}
  \bibinfo{year}{2019}\natexlab{}.
\newblock \bibinfo{title}{{The LLVM Compiler Infrastructure}}.
\newblock \bibinfo{howpublished}{\url{https://llvm.org}}.
\newblock


\bibitem[\protect\citeauthoryear{Maas, Love, Stefanov, Tiwari, Shi, Asanovic,
  Kubiatowicz, and Song}{Maas et~al\mbox{.}}{2013}]%
        {maas-ccs13}
\bibfield{author}{\bibinfo{person}{Martin Maas}, \bibinfo{person}{Eric Love},
  \bibinfo{person}{Emil Stefanov}, \bibinfo{person}{Mohit Tiwari},
  \bibinfo{person}{Elaine Shi}, \bibinfo{person}{Krste Asanovic},
  \bibinfo{person}{John Kubiatowicz}, {and} \bibinfo{person}{Dawn Song}.}
  \bibinfo{year}{2013}\natexlab{}.
\newblock \showarticletitle{{Phantom: Practical oblivious computation in a
  secure processor}}. In \bibinfo{booktitle}{\emph{ACM SIGSAC Conference on
  Computer and Communications Security}}.
\newblock


\bibitem[\protect\citeauthoryear{McCune, Li, Qu, Zhou, Datta, Gligor, and
  Perrig}{McCune et~al\mbox{.}}{2010}]%
        {trustvisor}
\bibfield{author}{\bibinfo{person}{Jonathan~M. McCune}, \bibinfo{person}{Yanlin
  Li}, \bibinfo{person}{Ning Qu}, \bibinfo{person}{Zongwei Zhou},
  \bibinfo{person}{Anupam Datta}, \bibinfo{person}{Virgil Gligor}, {and}
  \bibinfo{person}{Adrian Perrig}.} \bibinfo{year}{2010}\natexlab{}.
\newblock \showarticletitle{{TrustVisor:} Efficient {TCB} Reduction and
  Attestation}. In \bibinfo{booktitle}{\emph{IEEE Symposium on Security and
  Privacy}}.
\newblock


\bibitem[\protect\citeauthoryear{Moghimi, Eisenbarth, and Sunar}{Moghimi
  et~al\mbox{.}}{2018}]%
        {MemJam}
\bibfield{author}{\bibinfo{person}{Ahmad Moghimi}, \bibinfo{person}{Thomas
  Eisenbarth}, {and} \bibinfo{person}{Berk Sunar}.}
  \bibinfo{year}{2018}\natexlab{}.
\newblock \showarticletitle{MemJam: A False Dependency Attack Against
  Constant-Time Crypto Implementations in SGX}. In
  \bibinfo{booktitle}{\emph{Topics in Cryptology -- CT-RSA 2018}},
  \bibfield{editor}{\bibinfo{person}{Nigel~P. Smart}} (Ed.).
  \bibinfo{publisher}{Springer International Publishing}.
\newblock


\bibitem[\protect\citeauthoryear{Moghimi, Irazoqui, and Eisenbarth}{Moghimi
  et~al\mbox{.}}{2017}]%
        {MIE17}
\bibfield{author}{\bibinfo{person}{Ahmad Moghimi}, \bibinfo{person}{Gorka
  Irazoqui}, {and} \bibinfo{person}{Thomas Eisenbarth}.}
  \bibinfo{year}{2017}\natexlab{}.
\newblock \bibinfo{booktitle}{\emph{{CacheZoom: How SGX} Amplifies The Power of
  Cache Attacks}}.
\newblock \bibinfo{type}{{T}echnical {R}eport}.
  \bibinfo{institution}{arXiv:1703.06986 [cs.CR]}.
\newblock
\newblock
\shownote{\url{https://arxiv.org/abs/1703.06986}.}


\bibitem[\protect\citeauthoryear{Ohrimenko, Costa, Fournet, Gkantsidis,
  Kohlweiss, and Sharma}{Ohrimenko et~al\mbox{.}}{2015}]%
        {ohrimenko2015observing}
\bibfield{author}{\bibinfo{person}{Olga Ohrimenko}, \bibinfo{person}{Manuel
  Costa}, \bibinfo{person}{C{\'e}dric Fournet}, \bibinfo{person}{Christos
  Gkantsidis}, \bibinfo{person}{Markulf Kohlweiss}, {and}
  \bibinfo{person}{Divya Sharma}.} \bibinfo{year}{2015}\natexlab{}.
\newblock \showarticletitle{Observing and preventing leakage in {MapReduce}}.
  In \bibinfo{booktitle}{\emph{ACM SIGSAC Conference on Computer and
  Communications Security}}. ACM.
\newblock


\bibitem[\protect\citeauthoryear{Ohrimenko, Schuster, Fournet, Meht, Nowozin,
  Vaswani, and Costa}{Ohrimenko et~al\mbox{.}}{2016}]%
        {OSF+16}
\bibfield{author}{\bibinfo{person}{Olga Ohrimenko}, \bibinfo{person}{Felix
  Schuster}, \bibinfo{person}{Cedric Fournet}, \bibinfo{person}{Aasthaa Meht},
  \bibinfo{person}{Sebastian Nowozin}, \bibinfo{person}{Kapil Vaswani}, {and}
  \bibinfo{person}{Manuel Costa}.} \bibinfo{year}{2016}\natexlab{}.
\newblock \showarticletitle{Oblivious Multi-Party Machine Learning on Trusted
  Processors}. In \bibinfo{booktitle}{\emph{USENIX Security Symposium}}.
\newblock


\bibitem[\protect\citeauthoryear{Osvik, Shamir, and Tromer}{Osvik
  et~al\mbox{.}}{2006}]%
        {Osv2006}
\bibfield{author}{\bibinfo{person}{Dag~Arne Osvik}, \bibinfo{person}{Adi
  Shamir}, {and} \bibinfo{person}{Eran Tromer}.}
  \bibinfo{year}{2006}\natexlab{}.
\newblock \showarticletitle{Cache Attacks and Countermeasures: The Case of
  {AES}}. In \bibinfo{booktitle}{\emph{The Cryptographers' Track at the {RSA}
  Conference on Topics in Cryptology}}.
\newblock
\showISBNx{3-540-31033-9, 978-3-540-31033-4}


\bibitem[\protect\citeauthoryear{Page}{Page}{2005}]%
        {Page05partitionedcache}
\bibfield{author}{\bibinfo{person}{D. Page}.} \bibinfo{year}{2005}\natexlab{}.
\newblock \showarticletitle{Partitioned Cache Architecture as a Side-Channel
  Defence Mechanism}. In \bibinfo{booktitle}{\emph{{IACR Eprint} archive}}.
\newblock


\bibitem[\protect\citeauthoryear{Pappas, Polychronakis, and Keromytis}{Pappas
  et~al\mbox{.}}{2012}]%
        {PaPoKe2012}
\bibfield{author}{\bibinfo{person}{Vasilis Pappas}, \bibinfo{person}{Michalis
  Polychronakis}, {and} \bibinfo{person}{Angelos~D. Keromytis}.}
  \bibinfo{year}{2012}\natexlab{}.
\newblock \showarticletitle{Smashing the Gadgets: Hindering Return-Oriented
  Programming Using In-Place Code Randomization}. In
  \bibinfo{booktitle}{\emph{IEEE Symposium on Security and Privacy}}.
\newblock


\bibitem[\protect\citeauthoryear{{PaX Team}}{{PaX Team}}{[n.d.]}]%
        {PaX-ASLR}
\bibfield{author}{\bibinfo{person}{{PaX Team}}.}
  \bibinfo{year}{[n.d.]}\natexlab{}.
\newblock \bibinfo{title}{PaX address space layout randomization (ASLR)}.
\newblock
  \bibinfo{howpublished}{\url{http://pax.grsecurity.net/docs/aslr.txt}}.
\newblock


\bibitem[\protect\citeauthoryear{Portela, Barbosa, Scerri, Warinschi, Bahmani,
  Brasser, and Sadeghi}{Portela et~al\mbox{.}}{2017}]%
        {smpc-sgx}
\bibfield{author}{\bibinfo{person}{Bernardo Portela}, \bibinfo{person}{Manuel
  Barbosa}, \bibinfo{person}{Guillaume Scerri}, \bibinfo{person}{Bogdan
  Warinschi}, \bibinfo{person}{Raad Bahmani}, \bibinfo{person}{Ferdinand
  Brasser}, {and} \bibinfo{person}{Ahmad-Reza Sadeghi}.}
  \bibinfo{year}{2017}\natexlab{}.
\newblock \showarticletitle{Secure Multiparty Computation from SGX}. In
  \bibinfo{booktitle}{\emph{Financial Cryptography and Data Security}}.
\newblock


\bibitem[\protect\citeauthoryear{Rane, Lin, and Tiwari}{Rane
  et~al\mbox{.}}{2015}]%
        {RaLiTi2015}
\bibfield{author}{\bibinfo{person}{Ashay Rane}, \bibinfo{person}{Calvin Lin},
  {and} \bibinfo{person}{Mohit Tiwari}.} \bibinfo{year}{2015}\natexlab{}.
\newblock \showarticletitle{Raccoon: Closing Digital Side-channels Through
  Obfuscated Execution}. In \bibinfo{booktitle}{\emph{USENIX Security
  Symposium}}.
\newblock
\urldef\tempurl%
\url{http://dl.acm.org/citation.cfm?id=2831143.2831171}
\showURL{%
\tempurl}


\bibitem[\protect\citeauthoryear{Ren, Fletcher, Kwon, Stefanov, Shi, Van~Dijk,
  and Devadas}{Ren et~al\mbox{.}}{2015}]%
        {ren2015constants}
\bibfield{author}{\bibinfo{person}{Ling Ren}, \bibinfo{person}{Christopher~W
  Fletcher}, \bibinfo{person}{Albert Kwon}, \bibinfo{person}{Emil Stefanov},
  \bibinfo{person}{Elaine Shi}, \bibinfo{person}{Marten Van~Dijk}, {and}
  \bibinfo{person}{Srinivas Devadas}.} \bibinfo{year}{2015}\natexlab{}.
\newblock \showarticletitle{Constants Count: Practical Improvements to
  Oblivious {RAM}}. In \bibinfo{booktitle}{\emph{USENIX Security Symposium}}.
\newblock


\bibitem[\protect\citeauthoryear{Roemer, Buchanan, Shacham, and Savage}{Roemer
  et~al\mbox{.}}{2012}]%
        {roemer2012return}
\bibfield{author}{\bibinfo{person}{Ryan Roemer}, \bibinfo{person}{Erik
  Buchanan}, \bibinfo{person}{Hovav Shacham}, {and} \bibinfo{person}{Stefan
  Savage}.} \bibinfo{year}{2012}\natexlab{}.
\newblock \showarticletitle{Return-oriented programming: Systems, languages,
  and applications}.
\newblock \bibinfo{journal}{\emph{ACM Transactions on Information and System
  Security}} \bibinfo{volume}{15}, \bibinfo{number}{1} (\bibinfo{year}{2012}).
\newblock


\bibitem[\protect\citeauthoryear{Sasy, Gorbunov, and Fletcher}{Sasy
  et~al\mbox{.}}{2017}]%
        {sasyzerotrace}
\bibfield{author}{\bibinfo{person}{Sajin Sasy}, \bibinfo{person}{Sergey
  Gorbunov}, {and} \bibinfo{person}{Christopher Fletcher}.}
  \bibinfo{year}{2017}\natexlab{}.
\newblock \showarticletitle{{ZeroTrace: Oblivious} Memory Primitives from
  {Intel SGX}}.
\newblock \bibinfo{journal}{\emph{IACR Cryptology ` Archive}}
  \bibinfo{volume}{Report 2017/549} (\bibinfo{year}{2017}).
\newblock


\bibitem[\protect\citeauthoryear{Schuster, Costa, Fournet, Gkantsidis, Peinado,
  Mainar-Ruiz, and Russinovich}{Schuster et~al\mbox{.}}{2015}]%
        {vc3}
\bibfield{author}{\bibinfo{person}{Felix Schuster}, \bibinfo{person}{Manuel
  Costa}, \bibinfo{person}{Cedric Fournet}, \bibinfo{person}{Christos
  Gkantsidis}, \bibinfo{person}{Marcus Peinado}, \bibinfo{person}{Gloria
  Mainar-Ruiz}, {and} \bibinfo{person}{Mark Russinovich}.}
  \bibinfo{year}{2015}\natexlab{}.
\newblock \showarticletitle{{VC3: Trustworthy Data Analytics in the Cloud Using
  SGX}}.
\newblock


\bibitem[\protect\citeauthoryear{Schwarz, Weiser, Gruss, Maurice, and
  Mangard}{Schwarz et~al\mbox{.}}{2017}]%
        {SWG+17}
\bibfield{author}{\bibinfo{person}{Michael Schwarz}, \bibinfo{person}{Samuel
  Weiser}, \bibinfo{person}{Daniel Gruss}, \bibinfo{person}{Cl{\'e}mentine
  Maurice}, {and} \bibinfo{person}{Stefan Mangard}.}
  \bibinfo{year}{{2017}}\natexlab{}.
\newblock \showarticletitle{{Malware Guard Extension: Using SGX to Conceal
  Cache Attacks}}. In \bibinfo{booktitle}{\emph{{Detection of Intrusions and
  Malware, and Vulnerability Assessment}}}.
\newblock


\bibitem[\protect\citeauthoryear{Seo, Lee, Kim, Shih, Shin, Han, and Kim}{Seo
  et~al\mbox{.}}{2017}]%
        {SGXShield2017}
\bibfield{author}{\bibinfo{person}{Jaebaek Seo}, \bibinfo{person}{Byounyoung
  Lee}, \bibinfo{person}{Seongmin Kim}, \bibinfo{person}{Ming-Wei Shih},
  \bibinfo{person}{Insik Shin}, \bibinfo{person}{Dongsu Han}, {and}
  \bibinfo{person}{Taesoo Kim}.} \bibinfo{year}{2017}\natexlab{}.
\newblock \showarticletitle{{SGX-Shield}: Enabling Address Space Layout
  Randomization for {SGX} Programs}. In \bibinfo{booktitle}{\emph{Network and
  Distributed System Security Symposium}}.
\newblock


\bibitem[\protect\citeauthoryear{Shih, Lee, Kim, and Peinado}{Shih
  et~al\mbox{.}}{2017}]%
        {t-sgx}
\bibfield{author}{\bibinfo{person}{Ming-Wei Shih}, \bibinfo{person}{Sangho
  Lee}, \bibinfo{person}{Taesoo Kim}, {and} \bibinfo{person}{Marcus Peinado}.}
  \bibinfo{year}{2017}\natexlab{}.
\newblock \showarticletitle{{T-SGX: Eradicating} Controlled-Channel Attacks
  Against Enclave Programs}. In \bibinfo{booktitle}{\emph{Network and
  Distributed System Security Symposium}}.
\newblock


\bibitem[\protect\citeauthoryear{Sinha, Rajamani, and Seshia}{Sinha
  et~al\mbox{.}}{2017}]%
        {Sinha2017}
\bibfield{author}{\bibinfo{person}{Rohit Sinha}, \bibinfo{person}{Sriram
  Rajamani}, {and} \bibinfo{person}{Sanjit~A. Seshia}.}
  \bibinfo{year}{2017}\natexlab{}.
\newblock \showarticletitle{A compiler and verifier for page access oblivious
  computation}. In \bibinfo{booktitle}{\emph{Proceedings of the 2017 11th Joint
  Meeting on Foundations of Software Engineering - {ESEC}/{FSE} 2017}}.
  \bibinfo{publisher}{{ACM} Press}.
\newblock
\urldef\tempurl%
\url{https://doi.org/10.1145/3106237.3106248}
\showDOI{\tempurl}


\bibitem[\protect\citeauthoryear{Stecklina and Prescher}{Stecklina and
  Prescher}{2018}]%
        {LazyFP}
\bibfield{author}{\bibinfo{person}{Julian Stecklina} {and}
  \bibinfo{person}{Thomas Prescher}.} \bibinfo{year}{2018}\natexlab{}.
\newblock \showarticletitle{LazyFP: Leaking {FPU} Register State using
  Microarchitectural Side-Channels}.
\newblock \bibinfo{journal}{\emph{CoRR}}  \bibinfo{volume}{abs/1806.07480}
  (\bibinfo{year}{2018}).
\newblock
\showeprint[arxiv]{1806.07480}
\urldef\tempurl%
\url{http://arxiv.org/abs/1806.07480}
\showURL{%
\tempurl}


\bibitem[\protect\citeauthoryear{Stefanov, Van~Dijk, Shi, Fletcher, Ren, Yu,
  and Devadas}{Stefanov et~al\mbox{.}}{2013}]%
        {stefanov-ccs13}
\bibfield{author}{\bibinfo{person}{Emil Stefanov}, \bibinfo{person}{Marten
  Van~Dijk}, \bibinfo{person}{Elaine Shi}, \bibinfo{person}{Christopher
  Fletcher}, \bibinfo{person}{Ling Ren}, \bibinfo{person}{Xiangyao Yu}, {and}
  \bibinfo{person}{Srinivas Devadas}.} \bibinfo{year}{2013}\natexlab{}.
\newblock \showarticletitle{{Path ORAM: an extremely simple oblivious RAM
  protocol}}. In \bibinfo{booktitle}{\emph{ACM SIGSAC Conference on Computer
  and Communications Security}}.
\newblock


\bibitem[\protect\citeauthoryear{Tsai, Porter, and Vij}{Tsai
  et~al\mbox{.}}{2017}]%
        {graphene}
\bibfield{author}{\bibinfo{person}{{Chia-che} Tsai}, \bibinfo{person}{Donald~E.
  Porter}, {and} \bibinfo{person}{Mona Vij}.} \bibinfo{year}{2017}\natexlab{}.
\newblock \showarticletitle{{Graphene-SGX: A Practical Library OS for
  Unmodified Applications on SGX}}. In \bibinfo{booktitle}{\emph{USENIX Annual
  Technical Conference}}.
\newblock


\bibitem[\protect\citeauthoryear{Van~Bulck, Minkin, Weisse, Genkin, Kasikci,
  Piessens, Silberstein, Wenisch, Yarom, and Strackx}{Van~Bulck
  et~al\mbox{.}}{2018}]%
        {foreshadow}
\bibfield{author}{\bibinfo{person}{Jo Van~Bulck}, \bibinfo{person}{Marina
  Minkin}, \bibinfo{person}{Ofir Weisse}, \bibinfo{person}{Daniel Genkin},
  \bibinfo{person}{Baris Kasikci}, \bibinfo{person}{Frank Piessens},
  \bibinfo{person}{Mark Silberstein}, \bibinfo{person}{Thomas~F. Wenisch},
  \bibinfo{person}{Yuval Yarom}, {and} \bibinfo{person}{Raoul Strackx}.}
  \bibinfo{year}{2018}\natexlab{}.
\newblock \showarticletitle{{Foreshadow: Extracting the Keys to the Intel {SGX}
  Kingdom with Transient Out-of-Order Execution}}. In
  \bibinfo{booktitle}{\emph{27th {USENIX} Security Symposium}}.
\newblock


\bibitem[\protect\citeauthoryear{Van~Bulck, Weichbrodt, Kapitza, Piessens, and
  Strackx}{Van~Bulck et~al\mbox{.}}{2017}]%
        {Bulck2017}
\bibfield{author}{\bibinfo{person}{Jo Van~Bulck}, \bibinfo{person}{Nico
  Weichbrodt}, \bibinfo{person}{R{\"u}diger Kapitza}, \bibinfo{person}{Frank
  Piessens}, {and} \bibinfo{person}{Raoul Strackx}.}
  \bibinfo{year}{2017}\natexlab{}.
\newblock \showarticletitle{{Telling Your Secrets without Page Faults: Stealthy
  Page Table-Based Attacks on Enclaved Execution}}. In
  \bibinfo{booktitle}{\emph{26th {USENIX} Security Symposium}}.
\newblock


\bibitem[\protect\citeauthoryear{van Schaik, Giuffrida, Bos, and Razavi}{van
  Schaik et~al\mbox{.}}{2018}]%
        {SGB+18}
\bibfield{author}{\bibinfo{person}{Stephan van Schaik},
  \bibinfo{person}{Cristiano Giuffrida}, \bibinfo{person}{Herbert Bos}, {and}
  \bibinfo{person}{Kaveh Razavi}.} \bibinfo{year}{2018}\natexlab{}.
\newblock \showarticletitle{{Malicious Management Unit: Why Stopping Cache
  Attacks in Software is Harder Than You Think}}. In
  \bibinfo{booktitle}{\emph{27th {USENIX} Security Symposium}}.
\newblock
\showISBNx{978-1-931971-46-1}
\urldef\tempurl%
\url{https://www.usenix.org/conference/usenixsecurity18/presentation/van-schaik}
\showURL{%
\tempurl}


\bibitem[\protect\citeauthoryear{Wang and Lee}{Wang and Lee}{2008}]%
        {WangLee2008}
\bibfield{author}{\bibinfo{person}{Zhenghong Wang} {and}
  \bibinfo{person}{Ruby~B. Lee}.} \bibinfo{year}{2008}\natexlab{}.
\newblock \showarticletitle{A Novel Cache Architecture with Enhanced
  Performance and Security}. In \bibinfo{booktitle}{\emph{Annual IEEE/ACM
  International Symposium on Microarchitecture}}.
\newblock


\bibitem[\protect\citeauthoryear{Wartell, Mohan, Hamlen, and Lin}{Wartell
  et~al\mbox{.}}{2012}]%
        {WaMoHaLi2012}
\bibfield{author}{\bibinfo{person}{Richard Wartell}, \bibinfo{person}{Vishwath
  Mohan}, \bibinfo{person}{Kevin~W. Hamlen}, {and} \bibinfo{person}{Zhiqiang
  Lin}.} \bibinfo{year}{2012}\natexlab{}.
\newblock \showarticletitle{Binary Stirring: Self-randomizing Instruction
  Addresses of Legacy x86 Binary Code}. In \bibinfo{booktitle}{\emph{ACM SIGSAC
  Conference on Computer and Communications Security}}.
\newblock


\bibitem[\protect\citeauthoryear{Williams and Sion}{Williams and Sion}{2012}]%
        {williams2012round}
\bibfield{author}{\bibinfo{person}{Peter Williams} {and} \bibinfo{person}{Radu
  Sion}.} \bibinfo{year}{2012}\natexlab{}.
\newblock \showarticletitle{Round-optimal access privacy on outsourced
  storage}. In \bibinfo{booktitle}{\emph{ACM Conference on Computer and
  Communications Security}}.
\newblock


\bibitem[\protect\citeauthoryear{Xu, Cui, and Peinado}{Xu
  et~al\mbox{.}}{2015}]%
        {Xu2015}
\bibfield{author}{\bibinfo{person}{Yuanzhong Xu}, \bibinfo{person}{Weidong
  Cui}, {and} \bibinfo{person}{Marcus Peinado}.}
  \bibinfo{year}{2015}\natexlab{}.
\newblock \showarticletitle{Controlled-Channel Attacks: Deterministic Side
  Channels for Untrusted Operating Systems}. In \bibinfo{booktitle}{\emph{IEEE
  Symposium on Security and Privacy}}.
\newblock


\bibitem[\protect\citeauthoryear{Yarom, Genkin, and Heninger}{Yarom
  et~al\mbox{.}}{2016}]%
        {CacheBleed2016}
\bibfield{author}{\bibinfo{person}{Y. Yarom}, \bibinfo{person}{D. Genkin},
  {and} \bibinfo{person}{N. Heninger}.} \bibinfo{year}{2016}\natexlab{}.
\newblock \bibinfo{booktitle}{\emph{{CacheBleed: A timing attack on OpenSSL
  constant time RSA}}}.
\newblock \bibinfo{type}{{T}echnical {R}eport}.
  \bibinfo{institution}{Cryptology ePrint Archive. Report 2016/224}.
\newblock
\newblock
\shownote{\url{https://eprint.iacr.org/2016/224.pdf}.}


\end{thebibliography}

\end{document}